\documentclass{aa}  
\usepackage{graphicx}
\usepackage{txfonts}
\newcommand{\mlpd}[2]{\bgroup\markoverwith{\textcolor{red}{\rule[0.5ex]{2pt}{0.4pt}}}\ULon{#1} \textcolor{blue} {#2}}

\def\hdu18{HD\,189733\,b}
\def\hd20{HD\,209458\,b}
\def\gj34{GJ\,3470\,b}

\def\co2{CO$_2$}
\def\ch4{CH$_4$}
\def\h2{H$_{2}$}
\def\h2o{H$_2$O}
\def\hep{He$^{+}$}
\def\hes{He(1$^{1}$S)}
\def\het{He(2$^{3}$S)}
\def\he_rat{He\,{\sc i}$_{\lambda 10833}$/He\,{\sc i}$_{\lambda 10832}$}
\def\mlr{$\dot M$}
\def\lya{Ly$\alpha$}
\def\ha{H$\alpha$}

\def\kms{km\,s$^{-1}$}

\def\rp{$R_{\rm P}$}
\def\rj{$R_{\rm Jup}$\ }
\def\mj{$M_{\rm Jup}$\ }

\def\gs{g\,s$^{-1}$}
\def\ecs{erg\,cm$^{-2}$\,s$^{-1}$}

\begin{document} 

\title{Modelling the He {\sc i} triplet absorption at 10830\,\AA\ in the atmospheres of HD 189733 b and GJ 3470 b}
%
%\subtitle{}
%
\titlerunning{Modelling of \het \, in HD 189733 b and GJ 3470 b}
\author{
M.~Lamp{\'o}n\inst{1}, M.~L\'opez-Puertas\inst{1}, 
J.~Sanz-Forcada\inst{2}, 
A.~S\'anchez-L\'opez\inst{3}, 
K.~Molaverdikhani\inst{4,5}, 
S.~Czesla\inst{6}, 
A.~Quirrenbach\inst{5},
E.~Pall{\'e}\inst{7,8}, 
J.\,A.~Caballero\inst{2},
Th.~Henning\inst{4},
M.~Salz\inst{6},
L.~Nortmann\inst{9},
J.~Aceituno\inst{1,10},
P.\,J.~Amado\inst{1},
F.\,F.~ Bauer\inst{1},
D.~Montes\inst{11},
E.~Nagel\inst{12},  
A.~Reiners\inst{9}, and 
I.~Ribas\inst{13,14} }
\institute{Instituto de Astrof{\'i}sica de Andaluc{\'i}a (IAA-CSIC), Glorieta de la Astronom{\'i}a s/n, 18008 Granada, Spain\\
\email{mlampon@iaa.es}
\and
Centro de Astrobiolog{\'i}a (CSIC-INTA), ESAC, Camino bajo del castillo s/n, 28692 Villanueva de la Ca{\~n}ada, Madrid, Spain
\and
Leiden Observatory, Leiden University, Postbus 9513, 2300 RA, Leiden, The Netherlands
\and
Max-Planck-Institut f{\"u}r Astronomie, K{\"o}nigstuhl 17, 69117 Heidelberg, Germany
\and
Landessternwarte, Zentrum f\"ur Astronomie der Universit\"at Heidelberg, K\"onigstuhl 12, 69117 Heidelberg, Germany
\and
Hamburger Sternwarte, Universit{\"a}t Hamburg, Gojenbergsweg 112, 21029 Hamburg, Germany
\and
Instituto de Astrof{\'i}sica de Canarias (IAC), Calle V{\'i}a L{\'a}ctea s/n, 38200 La Laguna, Tenerife, Spain
\and
Departamento de Astrof{\'i}sica, Universidad de La Laguna, 38026  La Laguna, Tenerife, Spain
\and
Institut f{\"u}r Astrophysik, Georg-August-Universit{\"a}t, Friedrich-Hund-Platz 1, 37077 G{\"o}ttingen, Germany
\and
Observatorio de Calar Alto, Sierra de los Filabres, 04550 G\'ergal, Almer\'{\i}a, Spain
\and
Departamento de F{\'i}sica de la Tierra y Astrof{\'i}sica \& IPARCOS-UCM (Instituto de F{\'i}sica de Part{\'i}culas y del Cosmos de la UCM), Facultad de Ciencias F{\'i}sicas, Universidad Complutense de Madrid,  28040 Madrid, Spain
\and
Th{\"u}ringer Landessternwarte Tautenburg, Sternwarte 5, 07778 Tautenburg, Germany
\and
Institut de Ci\`encies de l'Espai (CSIC-IEEC), Campus UAB, c/ de Can Magrans s/n, 08193 Bellaterra, Barcelona, Spain
\and
Institut d'Estudis Espacials de Catalunya (IEEC), 08034 Barcelona, Spain
}
\authorrunning{M. Lampón et al.}
\date{Received 13 September 2020 / Accepted 20 January 2021}

\abstract
{Characterising the atmospheres of exoplanets is key to understanding their nature and provides hints about their formation and evolution.
High resolution measurements of the helium triplet absorption of highly irradiated planets have been recently reported, which provide a new means of studying their atmospheric escape.
In this work we study the escape of the upper atmospheres of \hdu18 and \gj34 by analysing high resolution He~{\sc i} triplet absorption measurements and using a 1D hydrodynamic spherically symmetric model coupled with a non-local thermodynamic model for the He~{\sc i} triplet state. We also use the H density derived from \lya\ observations to further constrain their temperatures, mass-loss rates, and H/He ratios.
We have significantly improved our knowledge of the upper atmospheres of these planets. 
While \hdu18 has a rather compressed atmosphere and small gas radial velocities, \gj34, on the other hand with a gravitational potential ten times smaller, exhibits a very extended atmosphere and large radial outflow velocities. Hence, although \gj34 is much less irradiated in the X-ray and extreme ultraviolet radiation, and its upper atmosphere is much cooler, it evaporates at a comparable rate.
In particular, we find that the upper atmosphere of \hdu18 is compact and hot, with a maximum temperature of 12\,400$^{+400}_{-300}$\,K, with a very low mean molecular mass 
(H/He=(99.2/0.8)$\pm0.1$), 
which is almost fully ionised above 1.1\,\rp, and with a mass-loss rate of (1.1$\pm0.1$)\,$\times$\,10$^{11}$\,\gs. 
In contrast, the upper atmosphere of \gj34 is highly extended and relatively cold, with a maximum temperature of 5100$\pm900$\,K, also with a very low mean molecular mass  (H/He=(98.5/1.5)$^{+1.0}_{-1.5}$), which is not strongly ionised, and with a mass-loss rate of (1.9$\pm1.1$)\,$\times$\,10$^{11}$\,\gs.
Furthermore, our results suggest that upper atmospheres of giant planets undergoing hydrodynamic escape tend to have a very low mean molecular mass (H/He\,$\gtrsim$\,97/3).
}


\maketitle
%
%-------------------------------------------------------------------

%%%%%%%%%%%%%%%%%%%%%%%%%%%%%%%%
\section{Introduction} \label{Intro}
%%%%%%%%%%%%%%%%%%%%%%%%%%%%%%%%

Observations of atmospheres undergoing hydrodynamic escape  
provide critical information about their physical properties and can also offer important hints about their formation and evolution \citep[e.g.][]{Baraffe_2004,Baraffe_2005,Owen_2013,Owen_2017,Lopez_2013,Garcia_munoz_2019}. Hydrodynamic atmospheric escape is the most efficient atmospheric process of mass loss
\cite[see e.g.][]{Watson_1981,Yelle_2004,Tian2005,Garcia_munoz_2007,Salz_2015}. 
This process occurs when the gas pressure gradient overcomes the gravity of the planet at some atmospheric altitude, as it is heated via photo-ionisation. Therefore, the stellar irradiation, mainly X-ray and extreme ultraviolet (XUV) radiation, as well as the near-ultraviolet (NUV) radiation in exoplanets orbiting hot stars \cite[][]{Garcia_munoz_2019}, triggers hydrodynamic atmospheric escape generating a strong wind that substantially expands the thermosphere of the planet and ejects the gas beyond the Roche lobe.

\lya\ observations can probe extended atmospheres and provide important information about the planetary upper atmosphere \citep{VidalMadjar2003}. 
However, \lya\ can only be observed from space.  
Moreover, geocoronal emission contamination and interstellar medium absorption dominate the core of the line, leaving only their wings with potential exoplanetary information \cite[see e.g.][]{VidalMadjar2003,Ehrenreich_2008}.
Observations of X-ray radiation and ultraviolet lines from heavy elements (e.g. O{\sc\,i} and C{\sc\,ii}) in exoplanets undergoing hydrodynamic atmospheric escape have similar limitations \citep[see e.g.][]{VidalMadjar2004,Ben_Jaffel_2013,Poppenhaeger_2013}.

Observations of the He\,{\sc i} 2$^3$S--2$^3$P lines\footnote{At wavelengths\,10832.06, 10833.22, and 10833.31\,\AA\ in the vacuum, often referred to as their air wavelengths of  10830\,\AA.}
(hereafter \het\, lines) and \ha\ are not seriously limited by interstellar absorption
\citep[see e.g.][]{Spake_2018,Nortmann2018,Allart_2018,Mansfield_2018,Allart_2019,Yan_2018,Casasayas_2018,Wyttenbach_2020}.
Therefore, detailed studies of the absorption line profile are feasible. For instance, \cite{Lampon2020} have recently analysed the \het\ absorption in the atmosphere of \hd20 and derived a well-constrained relationship between the mass-loss rate and temperature, as well as key atmospheric parameters such as the \het\ density, the [H]/[H$^+$] transition altitude, and the XUV absorption effective radii. 

In addition, as \het\ measurements probe different atmospheric altitudes than the H lines, it is possible to reduce the degeneracy significantly by combining information from both elements. 
Indeed, by comparing the hydrogen density profile retrieved from \lya\ by \cite{Salz2016} with that derived from \het\ observations, \cite{Lampon2020} found H/He\,$\approx$\,98/2 in the upper atmosphere of \hd20, which is significantly higher than the commonly used value of 90/10 \citep[e.g.][]{Oklopcic2018, Mansfield_2018,Ninan_2020,guilluy_2020}.
Moreover, by constraining the upper atmospheric H/He ratio, we can gain important insights on the formation, evolution, and nature of the planet \cite[see, e.g.][]{Hu_2015,Malsky_2020}.
Therefore, it is important to measure the H/He ratio in other evaporating atmospheres.

The exoplanets \hdu18 and \gj34 undergo hydrodynamic escape, as was probed by the \lya\ and oxygen observations in \hdu18 \citep{Lecavelier_des_Etangs_2010,Lecavelier_des_Etangs_2012,Bourrier_2013a,Ben_Jaffel_2013}, and by \lya\ and \het\ observations for \gj34 \citep{Bourrier2018,Palle2020,Ninan_2020}.
These planets have rather different physical properties.
\cite{Salz2016} estimated that \hdu18, with a high gravitational potential, has a hot thermosphere with weak winds, whereas \gj34, with lower gravitational potential, has a relatively cool atmosphere with strong winds. 
Both exoplanets are also rather different from \hd20, as they have distinct bulk parameters and are irradiated at different XUV fluxes. To date, \het\ spectral absorption has been observed in \hdu18 by \cite{Salz2018} and \cite{guilluy_2020}, as well as in \gj34 by \cite{Palle2020} and  \cite{Ninan_2020}. 

In this work we aim to improve the characterisation of the upper atmospheres of \hdu18 and \gj34 from the \het\ spectral absorption measurements taken with the high–resolution spectrograph CARMENES
\footnote{Calar  Alto  high-Resolution  search  for M dwarfs with Exoearths with Near-infrared and optical Échelle Spectrographs, at the 3.5 m Calar Alto Telescope.}
\citep{Quirrenbach16,Quirrenbach18} 
by \cite{Salz2018} and \cite{Palle2020}, respectively. To that end, we applied a 1D hydrodynamic model with spherical symmetry together with an \het\ non-local thermodynamic equilibrium (non-LTE) model to calculate the \het\ concentration and gas radial velocity distributions. Subsequently, we used a high-resolution radiative transfer model for calculating the synthetic spectra as observed by CARMENES and compared them with the measurements. 
By exploring a wide range of input parameters, we derived constraints on the mass-loss rate, temperature, H/He composition, \het\ density, [H]/[H$^+$] transition altitude, and XUV absorption effective radii. Finally, we compared our H densities with those retrieved from \lya\ measurements in previous studies in order to constrain the H/He composition.
 
The paper is organised as follows. Sect.\,\ref{observations} summarises the \het\ observations of \hdu18 and \gj34; Sect.\,\ref{modelling} briefly describes the modelling of the \het\ density, the gas radial velocities, and the \het\ absorption; Sect.\,\ref{results} shows and discusses the results obtained; in Sect.\,\ref{comparison_prev} we compare temperatures and mass-loss rates with previous works; and in Sect.\,\ref{conclusions} we present a summary and our conclusions.

%%%%%%%%%%%%%%%%%%%%%%%%%%%%%%%%
\section{Observations of He\,{\sc i} triplet absorption of \hdu18 and \gj34}
\label{observations}
%%%%%%%%%%%%%%%%%%%%%%%%%%%%%%%%

The \het\ absorption profiles analysed here were observed in the atmospheres of \hdu18 and \gj34 by \cite{Salz2018} and \cite{Palle2020}, respectively, with the high-resolution spectrograph CARMENES.
We briefly summarise these observations below.

\hdu18\ shows a significant He excess absorption at mid-transit, with a mean absorption level of 0.88\,$\pm$\,0.04$\%$, and of 0.24$\pm$0.12\% and 0.46$\pm$0.06\% for the ingress and egress phases, respectively. The absorption (in the planetary rest frame) appears shifted to blue wavelengths by $-3.5\,\pm$\,0.4\,\kms\ and $-12.6\,\pm$\,1.0\,\kms\ during the mid-transit and egress, respectively, and it appears shifted to red wavelengths by 6.5\,$\pm$\,3.1\,\kms\ during the ingress. We caution, however, as \cite{Salz2018} remarked, that these velocities could be potentially affected by stellar pseudo-signals. 
Another important feature is that the ratio between the absorption in the stronger \het\ line, caused by the two unresolved lines centred at 10833.22 and 10833.31\,\AA, and the weaker one centred at 10832.06\,\AA \ (hereafter 
\he_rat) is 2.8\,$\pm$\,0.2, which is much smaller than expected from optically thin conditions.

\gj34 shows a 1.5\,$\pm$\,0.3\% He absorption depth at mid-transit \citep{Palle2020}. Unfortunately, the individual spectra during ingress or egress do not have a sufficient signal-to-noise ratio (S/N) to probe for any blue or red shifts. 
The mid-transit spectrum appears shifted to blue wavelengths by $-3.2\,\pm$\,1.3\,\kms. 
The analysis of these absorption profiles is discussed in Sect.~\ref{tran_spectra}.

%%%%%%%%%%%%%%%%%%%%%%%%%%%%%%%%%%%%%%%%%%%%%%%%%%%%%%%%%%%%%%%%
\section{Modelling the He\,{\sc i} triplet} \label{modelling}
%%%%%%%%%%%%%%%%%%%%%%%%%%%%%%%%%%%%%%%%%%%%%%%%%%%%%%%%%%%%%%%%%

%%%%%%%%%%%%%%%%%%%%%%%%%%%%%%%%%%%%%%%%%%%%%%%%%%%
\subsection{Helium triplet density} \label{modelling_density}
%%%%%%%%%%%%%%%%%%%%%%%%%%%%%%%%%%%%%%%%%%%%%%%%%%%

We calculated the populations of the \het\ by using the model described in \cite{Lampon2020}. 
Briefly, we used a 1D hydrodynamic  model together with a non-LTE model to calculate the \het\ density distribution in the substellar direction (the one that connects the star-planet centres) 
in the upper atmosphere of the planets. 
We assumed that the substellar conditions are a representative of the whole planetary sphere, so that a spherical symmetry was adopted.
The mass-loss rate derived under this assumption is a valid estimate for the whole atmosphere when divided by a factor of $\sim$4 to account for the 3D asymmetric stellar irradiation on the planetary surface \cite[see e.g.][]{Murray-Clay2009,Stone_2009,Tripathi_2015,Salz2016}.

The hydrodynamic equations were solved assuming that the escaping gas has a constant speed of sound, $v_s$\,=\,$\sqrt{k\,T(r)/\mu (r)}$, where $k$ is the Boltzmann constant, $T(r)$ is temperature, and $\mu (r)$ is the mean molecular weight. 
This assumption leads to the same analytical solution as the isothermal Parker wind solution. However, the atmosphere is not assumed to be isothermal, but the temperature varies with altitude in such a way that the $T(r)/\mu (r)$ ratio is constant. That is to say we assume 
$v_s$\,=\,$\sqrt{k\,T_0/\bar{\mu}}$ where $\bar{\mu}$ is the average mean molecular weight, calculated in the model, and $T_0$ is a model input parameter which is very similar to the maximum of the thermospheric temperature profile calculated by hydrodynamic models that solve the energy balance equation \citep[see, e.g.][]{Salz2016}.
The temperature ,$T_0$, the mass-loss rate, \mlr\ (of all species considered in the model), and the H/He mole-fraction ratio (i.e. the composition of the upper atmosphere) are input parameters to the model. 
The physical parameters of the planets, such as their mass, $M_{\rm P}$, and size, \rp, introduced in the model, are listed in Table\,\ref{table.parameters}.

\begin{table}
\centering
\caption{\label{table.parameters}System parameters of HD\,189733 and GJ\,3470.}
\begin{tabular}{l c l} 
\hline  \hline  \noalign{\smallskip}
Parameter & Value & Reference  \\
\noalign{\smallskip} \hline \noalign{\smallskip}
 \multicolumn{3}{l}{\em HD\,189733} \\
   \noalign{\smallskip}
$d$ &   
19.775\,$\pm$\,0.013\,pc & Gaia DR2$^{(a)}$\\
\noalign{\smallskip}
$R_{\star}$ &   
0.805\,$\pm$\,0.016\,$R_\sun$ & \citet{Boyajian_2015}\\
\noalign{\smallskip}
$M_{\star}$ & 0.846\,\,$^{+0.06}_{-0.049}$\,$M_\sun$    & \citet{de_Kok_2013}   \\
\noalign{\smallskip}
$T_{\rm eff}$ & 4875\,$\pm$\,43\,K      & \citet{Boyajian_2015} \\
\noalign{\smallskip}
$[\rm Fe/H]_{\star}$ & -0.03\,$\pm$\,0.05       & \citet{Bouchy_2005}   \\
\noalign{\smallskip}
$a$$^{(b)}$     & 0.0332\,$\pm$\,0.0010\,au     & \citet{Agol_2010} \\
\noalign{\smallskip}
$R_{\rm P}^{(c)}$ & 1.23\,$\pm$\,0.03\,\rj\     & \citet{Baluev_2015}   \\
\noalign{\smallskip}
$M_{\rm P}$     & 1.162\,$^{+0.058}_{-0.039}$\,\mj\     & \citet{de_Kok_2013}   \\
\noalign{\smallskip}
 \multicolumn{3}{l}{\em GJ\,3470} \\
   \noalign{\smallskip}
$d$ &   
29.45\,$\pm$\,0.05\,pc & Gaia DR2$^{(a)}$\\
\noalign{\smallskip}   
$R_{\star}$ &   
0.474\,$\pm$\,0.014\,\,$R_\sun$  & \citet{Palle2020}    \\
\noalign{\smallskip}
$M_{\star}$ & 0.476\,$\pm$\,0.019\,$M_\sun$     & \citet{Palle2020}     \\
\noalign{\smallskip}
$T_{\rm eff}$ & 3725\,$\pm$\,54\,K      & \citet{Palle2020}     \\
\noalign{\smallskip}
$[\rm Fe/H]_{\star}$ & +0.420\,$\pm$\,0.019     & \citet{Palle2020}     \\
\noalign{\smallskip}
$a$             & 0.0348\,$\pm$\,0.0014\,au     & \citet{Bonfils_2012} \\
\noalign{\smallskip}
$R_{\rm P}$     & 0.36\,$\pm$\,0.01\,\rj\       & \citet{Palle2020}     \\
\noalign{\smallskip}
$M_{\rm P}$     & 0.036\,$\pm$\,0.002\,\mj\     & \citet{Palle2020}     \\
\noalign{\smallskip}
\hline
\end{tabular}
\tablefoot{
\tablefoottext{a}{\cite{Gaia_2018}.}
\tablefoottext{b}{From a/$R_{\star}$ = 8.863(20) by \cite{Agol_2010} and $R_{\star}$ by \cite{Boyajian_2015}.}
\tablefoottext{c}{ From $R_{P}$/$R_{\star}$ = 0.15712(40) by \citet{Baluev_2015} and $R_{\star}$ by \cite{Boyajian_2015}.
}}
\end{table}

The model computes the radial distribution of the concentrations of the following species: neutral and ionised hydrogen, H$^0$ and H$^+$, respectively, as well as helium singlet and ionised helium, \hes\ and \hep, and \het. In addition, it also calculates the radial velocity of the gas. The production and loss terms and the corresponding rates are listed in Table 2 of \cite{Lampon2020}. They represent a minor extension of those considered by \cite{Oklopcic2018}, where two additional processes were included: the charge exchange reactions, $Q_{\rm He}$ and $Q_{{\rm He}^+}$, from \cite{Koskinen2013a}.
Other parameters such as the H, \hes, and \het\ photo-ionisation cross sections were taken as in \cite{Lampon2020}.

In the model we established the lower boundary conditions where hydrodynamic escape originates. This is usually assumed to occur at 
$\mu$bar--nbar levels \citep[see, e.g.][]{Garcia_munoz_2007,Koskinen2013a,Salz2016,Murray-Clay2009}, although its geometric altitude is uncertain as it depends on the pressure, temperature, and composition below, which are normally unknown. 
In Sects.\,\ref{ana_hd189} and \ref{ana_gj34},  we discuss the effects of the lower boundary conditions on the absorption profiles of these planets. Nominally, we assumed that hydrodynamic escape originates at 1.02\,\rp\ (slightly higher than the optical radius) with a density of 10$^{14}$ cm$^{-3}$.
The density at the lower boundary was chosen large enough so that the XUV radiation is fully absorbed by the atmosphere above, but it is consistent with the values given by the hydrostatic models below \cite[see, e.g.][]{Salz2016}.

%----------------------------------  Fig.
\begin{figure}[tbp]
  \centering
  \vspace{-0.5cm}
  \includegraphics[width=0.45\textwidth]{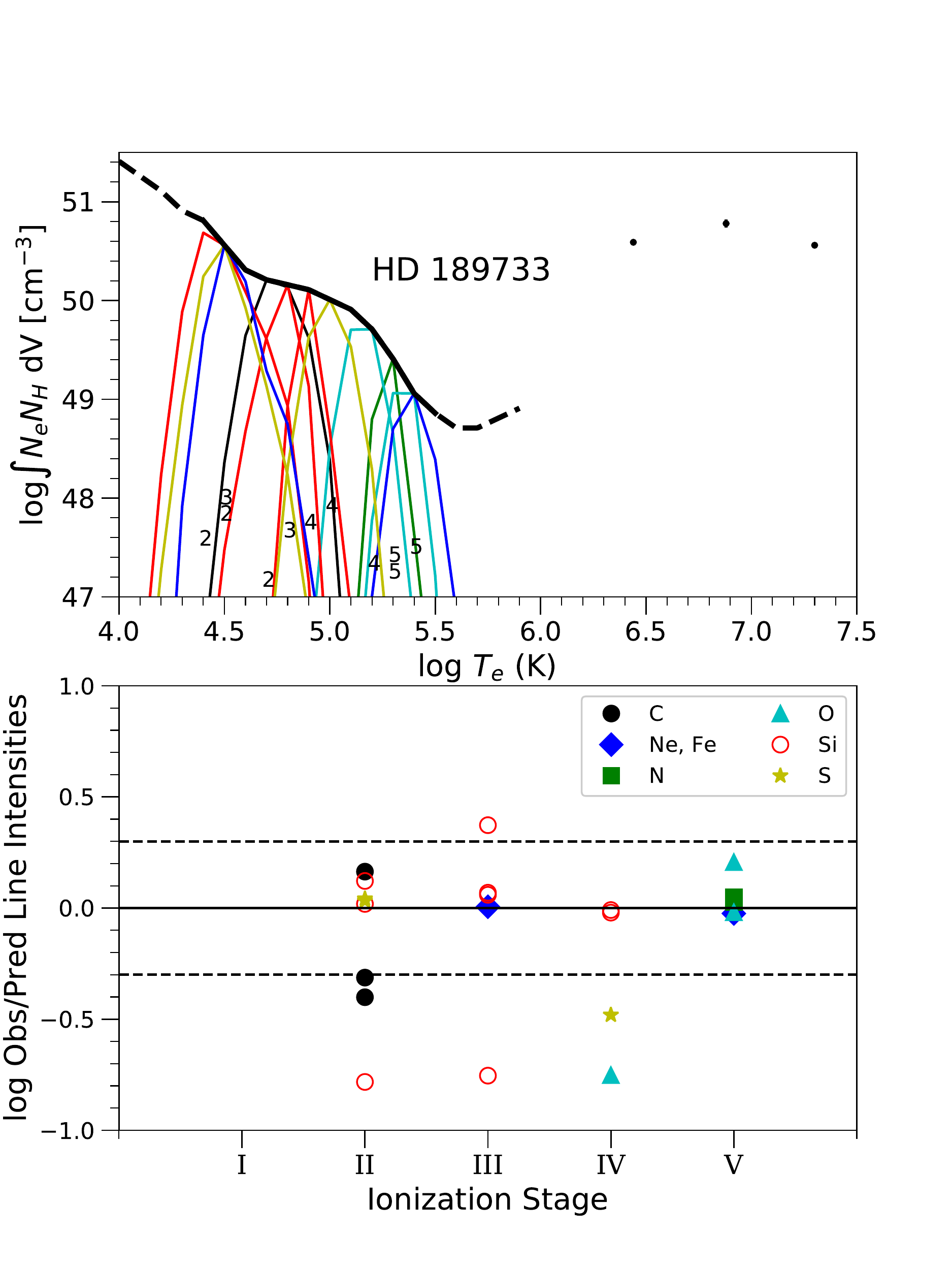}
  \caption{{\em Upper panel}: Emission measure distribution of
    HD\,189733 calculated using the line fluxes measured in the HSLA/COS summed spectrum. The 3-T model used to fit {\it XMM-Newton} summed EPIC spectra is also displayed. The thin lines represent the relative contribution function for each ion (the emissivity function multiplied by the EMD at each point), following same colour code as in lower panel. The small numbers indicate the ionisation stages of the species. {\em Lower panel}: Observed-to-predicted line flux ratios for the ion stages in the upper panel. The dotted lines denote a factor of 2.}
  \label{fig:emdhd189}
\end{figure}
%----------------------------------------------

\begin{figure}[htbp]
\includegraphics[angle=0.0,width=1\columnwidth]{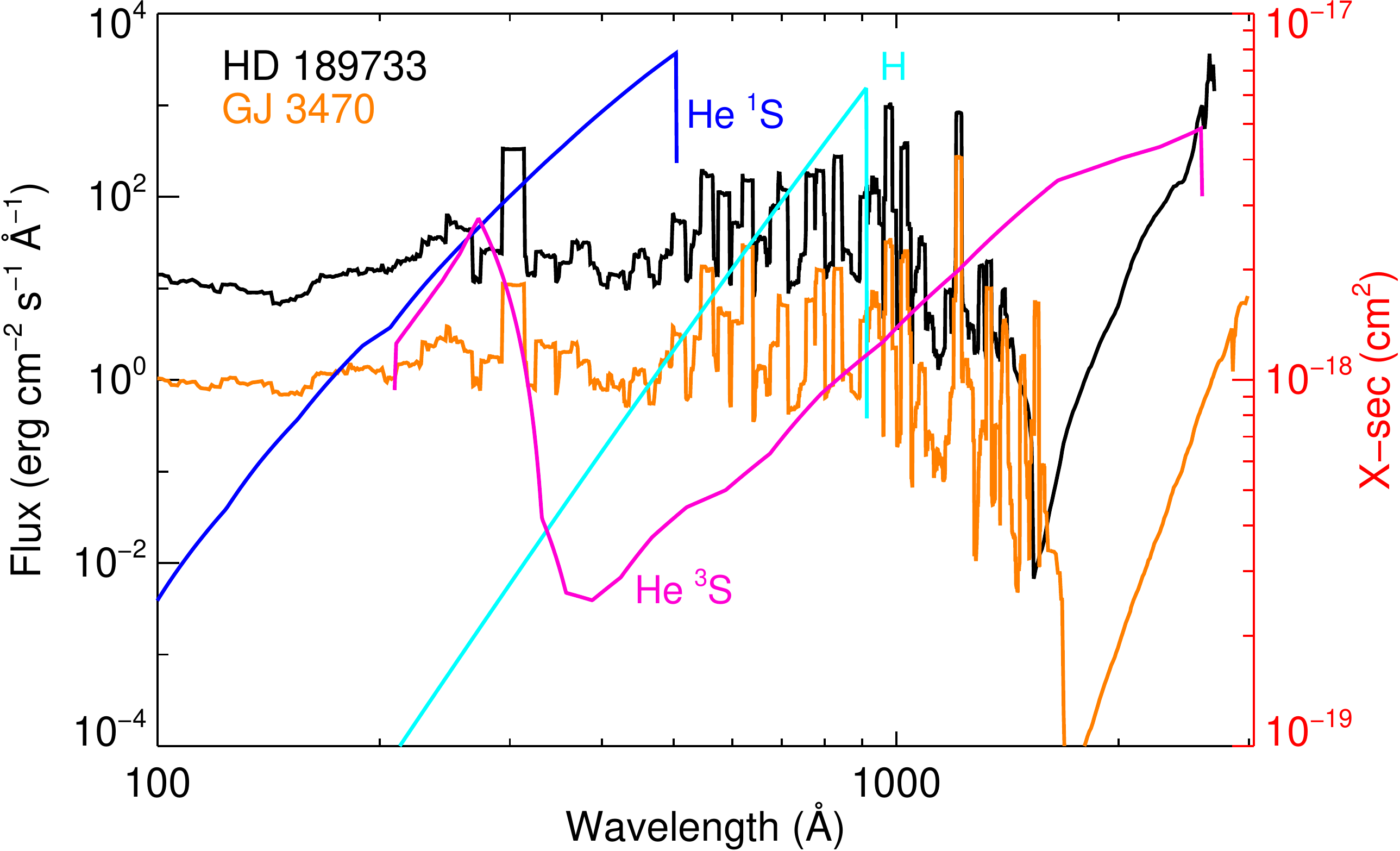}\hspace*{0.35cm}
\caption{Flux density (left y-axis) for HD\,189733 
 (black) and GJ\,3470 (orange) at 0.0332\,au and 0.0348\,au plotted at a resolution of 10\,\AA, respectively. The H, He singlet, and He triplet ionisation cross sections (right y-axis) are also shown.} 
\label{flux_xsec} 
\end{figure}

%%%%%%%%%%%%%%%%%%%%%%%%%%%%%%%%%%%%%%%%%%%%%%%%%%%%%%
\subsection{Stellar fluxes}\label{fluxes}
%%%%%%%%%%%%%%%%%%%%%%%%%%%%%%%%%%%%%%%%%%%%%%%%%%%%%%

A further input parameter required by the model is the stellar XUV spectral flux. 
For HD\,189733, a coronal model was used to obtain the stellar emission in the range of 5--1200~\AA. The model is based on the addition of all the {\it XMM-Newton} observations available to date of this star, between 2007 and 2015\,\footnote{Proposals id. 50607, 60097, 67239, 69089, 69229, 74498, 74839.}. 
The EPIC spectra were combined for a total exposure time of 461\,ks (pn), 822\,ks (MOS1), and 896\,ks (MOS2). The coronal 3-Temperature model ($\log~T$(K)$=6.44\pm0.01$, $6.88\pm0.01$, $7.3\pm0.01$, $\log EM$(cm$^{-3}$)$=50.59\pm0.03$, $50.78\pm0.04$, $50.56\pm0.01$, $L_{\rm X}=2.2\times10^{28}$\,erg\,s$^{-1}$) is complemented with line fluxes (Table~\ref{tab:cosfluxes}) from the HST/COS FUV spectrum available from the Hubble Spectral Legacy Archive (HSLA) to extend the model (Tables~\ref{tabemd} and \ref{tababund}) towards lower temperatures ($\log T$(K)$\sim4.0-5.9$), following \citet{san11}. This model (Fig.~\ref{fig:emdhd189}) is a substantial improvement with respect to the X-exoplanets model available in \citet{san11}. The modelled spectral energy distribution (SED) fluxes now indicate an EUV luminosity of $L_{\rm EUV H}=1.6\times10^{29}$\,erg\,s$^{-1}$ and $L_{\rm EUV He}=5.1\times10^{28}$\,erg\,s$^{-1}$ in the ranges of 100--912~\AA\ and 100-504~\AA, respectively. 
The SED covers the range of 5--1145~\AA, and it was generated following \citet{san11}. The data include several small flares, which were not removed on purpose, in order to provide an average model of active and non-active stages. We, therefore, used the summed HST/COS spectrum in the range of 1145--1450~\AA. The SED calculated using our model is consistent with the flux level observed in the actual {\it HST} observations of the star in the region of 1145$-$1200\,\AA.

{\it XMM-Newton} observations of GJ 3470 were used to model the corona of this  star, complemented with HST/STIS spectral line fluxes, as described in \citet{Bourrier2018}. The quality of the HST/COS spectra was not good enough to fix the UV continuum in the spectral range of 1150--1750~\AA, due to poor statistics, while HST/STIS coverage was limited to 1195--1248~\AA\ only. Thus, the model SED was also used in this spectral range.

In order to extend the SED of both planets to 2600\,\AA, we used the stellar atmospheric model of \cite{CK2004} scaled to the corresponding temperature, surface gravity, and metallicity (see Table\,\ref{table.parameters}). 
The composite SED for the spectral range of 5$-$2600\,\AA\ for both planets at their respective  orbital separations are shown in Fig.\,\ref{flux_xsec}.

\begin{figure*}[htbp]
\includegraphics[angle=0.0, width=1.0\columnwidth]{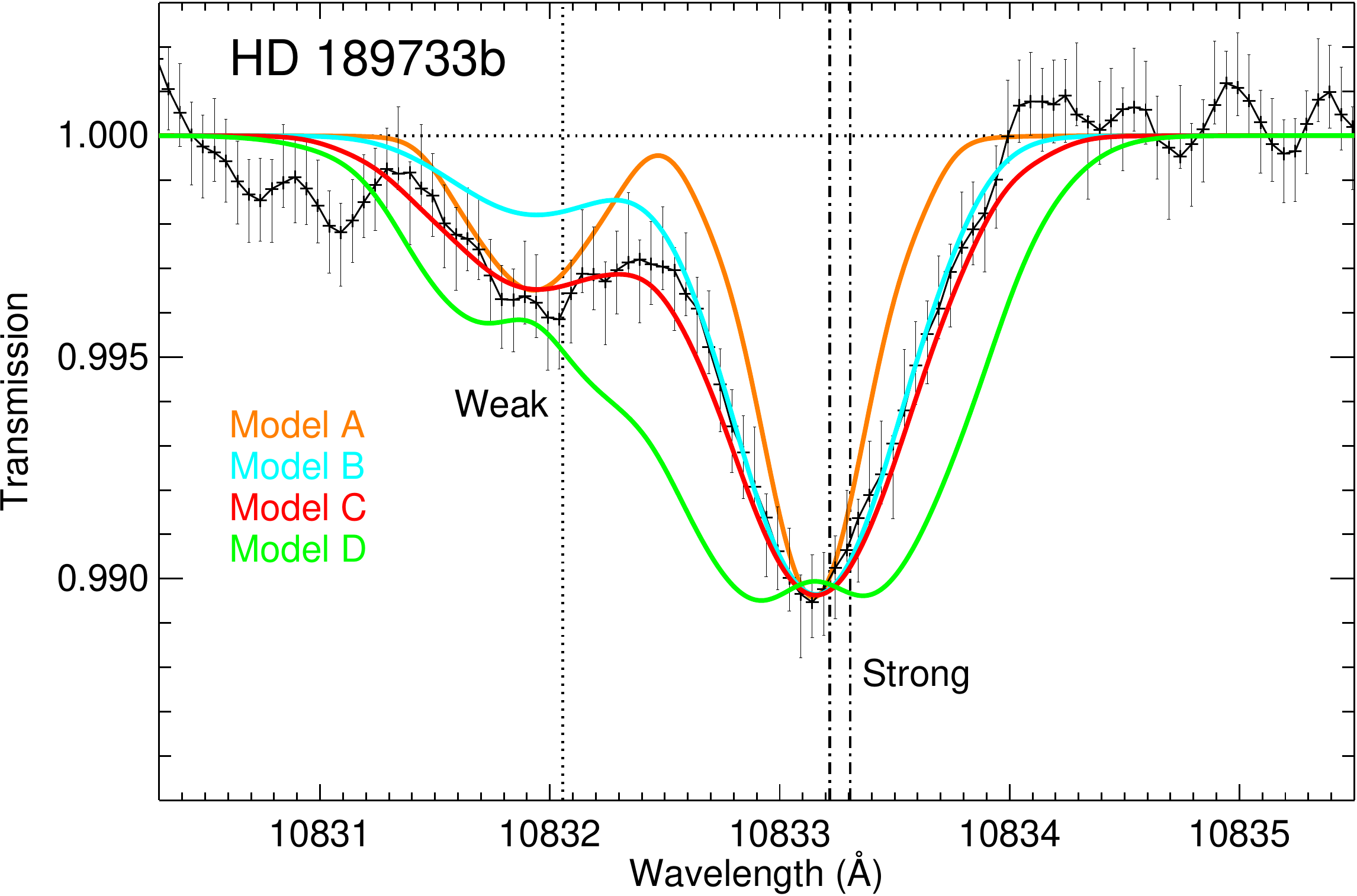}
\includegraphics[angle=0.0, width=1.0\columnwidth]{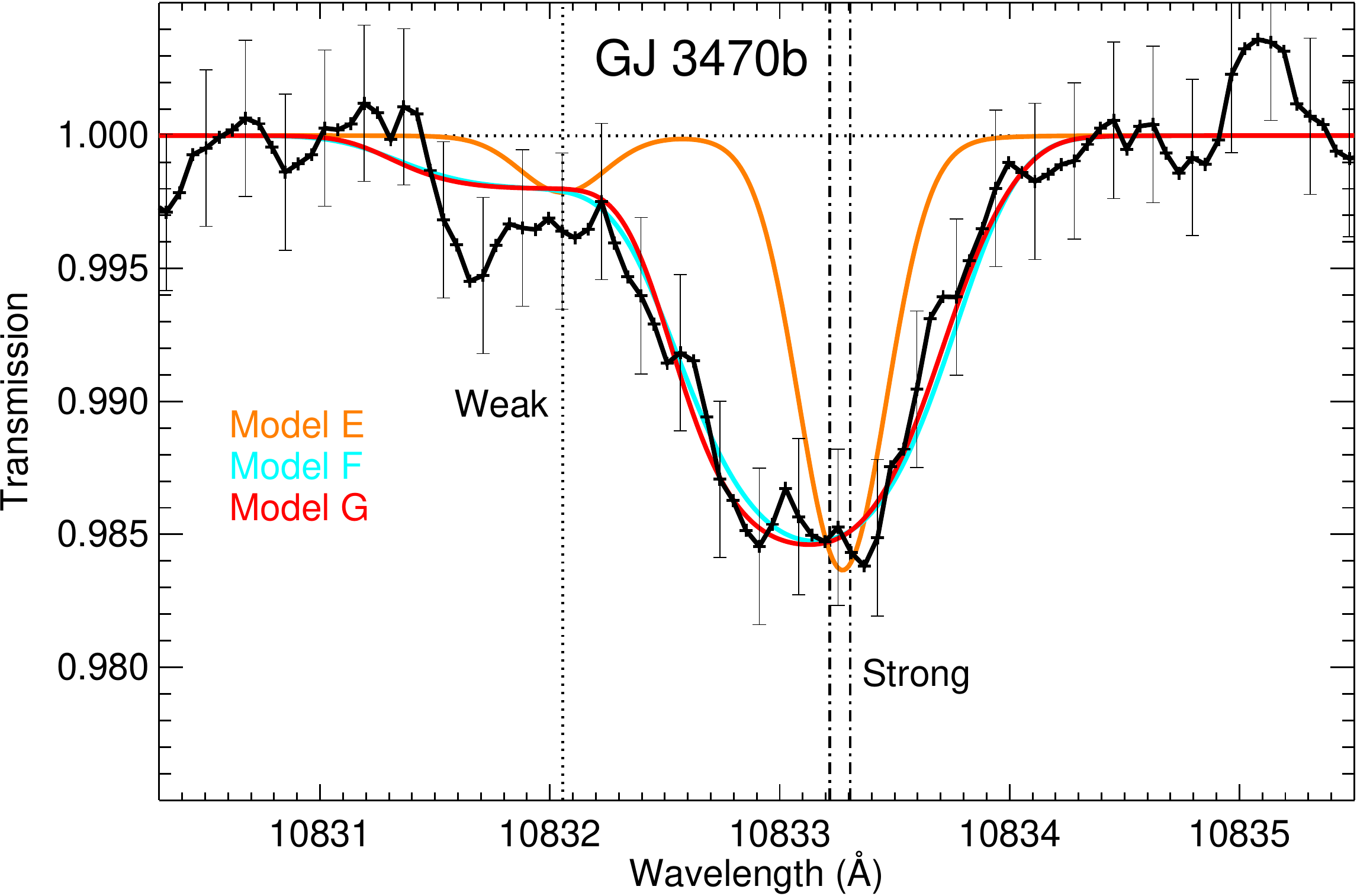} 
\caption{Spectral transmission of the He triplet at mid-transit for \hdu18 and \gj34\ (we note the different y-axis scale). Data points and their respective error bars are shown in black (adapted from \cite{Salz2018} and \cite{Palle2020}, respectively). Wavelengths are given in vacuum. The best-fit simulations are shown with red curves. For \hdu18, the best fit corresponds to a temperature of 10\,000\,K, a mass-loss rate of 10$^{9}$\,\gs\ and an H/He mole-fraction of 90/10. For \gj34, the best fit corresponds to a temperature of 6000\,K, $\dot{M}$=3.2$\times\,10^{10}$\,\gs\ and an H/He of 90/10. Other models are described in Table\,\ref{he_hd189_models}. 
The positions of the helium lines are marked by vertical  dotted (weak) and dot-dashed (strong) lines.}  
\label{absorption}
\end{figure*}

\begin{table*}[htbp]
\centering
\caption{\label{he_hd189_models}Models used for the calculation of the absorption shown in Fig.\,\ref{absorption}. }
\begin{tabular}{l c c c c} 
\hline  \hline  \noalign{\smallskip}
Model & Turbulent & LOS components & Lower  & Gas radial   \\
     &  broadening$^a$ &  &boundary layer &  velocity  \\
\noalign{\smallskip} \hline \noalign{\smallskip}
\multicolumn{5}{l}{\em \hdu18} \\
\noalign{\smallskip}
A &  No & --3.5\,\kms\ (100\%) & Yes  & This model$^b$ \\
B &  Yes & --12\,\kms (25\%), --3.5\,\kms (47\%), 5\,\kms (28\%),  & No  & This model$^b$ \\
C &  Yes & --12\,\kms (25\%), --3.5\,\kms (47\%), 5\,\kms (28\%),  & Yes  & This model$^b$ \\
D &  No & --3.5\,\kms\ (100\%) & Yes  & 40\,\kms$^{\, c}$ \\
\multicolumn{5}{l}{\em \gj34} \\
\noalign{\smallskip}
E &  Yes & No component & Yes  & No velocities \\
F &  Yes & --3.2\,\kms\, $\forall z$ & Yes  & This model \\
G &  Yes & No component at $z<R_{\rm lobe}$, --5\,\kms\,  at $z>R_{\rm lobe}$ & Yes  & This model \\
\noalign{\smallskip}
\hline
\end{tabular}
\tablefoot{
\tablefoottext{a}{Doppler broadening was included in all models.}
\tablefoottext{b}{From the hydrodynamic model of this work. The gas radial velocities for this planet}  have negligible effects.
\tablefoottext{c}{As in \cite{Seidel2020}.}
}
\end{table*}

%%%%%%%%%%%%%%%%%%%%%%%%%%%%%%%%%%%%%%%%%%%%%%%%%%%
\subsection{Spectral absorption} \label{modelling_absorption}
%%%%%%%%%%%%%%%%%%%%%%%%%%%%%%%%%%%%%%%%%%%%%%%%%%%

With the \het\ calculations and gas radial velocities from the model described above, we computed the \het\ absorption by using a radiative transfer code for the primary transit geometry \citep{Lampon2020}. The spectroscopic data for the three metastable helium lines were taken from the NIST Atomic Spectra Database\footnote{\tt https://www.nist.gov/pml/atomic-spectra-database.}.
Doppler line shapes are assumed at the atmospheric temperature used in the model density, and an additional broadening produced by turbulent velocities can be included if necessary \cite[$v_{\rm turb}$\,=\,$\sqrt{5kT/3m}$, 
where $m$ is the mass of an He atom, see Eq.\,16 in][]{Lampon2020}.
The component of the radial velocity of the gas along the line of sight (LOS) towards the observer is also included in order to account for the motion of \het\ as predicted in the hydrodynamic model. In addition to the radial velocities, 
averaged winds (e.g. day-to-night and super-rotation winds), and planetary rotation  \citep[see e.g.][]{Salz2018,Seidel2020} can also be included in the radiative transfer model, as required \cite[see Eq.\,15 in][]{Lampon2020}.

In this study, we performed the integration of the \het\ absorption up to 10\,\rp. This is motivated because we found that the \het\ distribution of \gj34\ is rather extended (see Sect.\,\ref{other_constrains}).

%%%%%%%%%%%%%%%%%%%%%%%%%%%%%%%%%%%%%%%%%%%%%%%%%%%%%%%%%%%%%%%%%
\subsection{Grid of simulations}\label{model_grid}
%%%%%%%%%%%%%%%%%%%%%%%%%%%%%%%%%%%%%%%%%%%%%%%%%%%%%%%%%%%%%%%%%

Here we have analysed the mid-transit absorption profiles of \hdu18 and \gj34 (see Fig.\,\ref{absorption}) in a similar manner as in \cite{Lampon2020} for \hd20. Briefly, from this model and the measured \het\ absorption, we could not unambiguously determine the mass-loss rate and temperature of the planetary atmosphere as these two quantities are degenerate in most cases. Thus, 
for a given H/He composition, we ran the \het\ model for a range of temperatures and mass-loss rates and computed the \het\ absorption. 
As also shown by \cite{Lampon2020}, the temperatures and mass-loss rates are also degenerate with respect to the H/He atomic ratio. We broke that degeneracy by also fitting the H$^0$ density profiles of the model to those derived from \lya\ measurements. To that end, we ran several sets of models for H/He atomic ratios ranging from 90/10, our nominal case, to 99.9/0.1 and to 99/1 for \hdu18 and \gj34, respectively.

Synthetic spectra from these simulations were
compared to the measured absorption profiles, and the corresponding reduced $\chi^2$ values were computed by \cite[see, e.g.][]{NR2007} 
\begin{equation*}
    \chi^2_{R}=\frac{\chi^2}{\nu} {\rm~~~with~~~}   \chi^2=\sum_i^N\frac{(Tr_{\rm mea,i}-Tr_{\rm mod,i})^2}{\sigma_i^2}, 
\end{equation*}
where ${\nu}=N-2$ is the number of degrees of freedom, $N$ is the number of fitted spectral points, $Tr_{\rm mea,i}$ and $Tr_{\rm mod,i}$ are the measured and calculated transmissions, and $\sigma_i$ is the error of the transmissions (see Fig.\,\ref{absorption}).
For obtaining the uncertainties in the derived parameters T and \mlr\ with this method, we considered the 95\% confidence levels of the $\chi^2$ (not the reduced $\chi^2$). In addition, we also explored the posterior probability distribution of the three model parameters for the grid of model spectra discussed above by using the Markov chain Monte Carlo (MCMC) method (Sect.~\ref{mcmc}).

%%%%%%%%%%%%%%%%%%%%%%%%%%%%%%%%
\section{Results and discussion} \label{results}
%%%%%%%%%%%%%%%%%%%%%%%%%%%%%%%%

%%%%%%%%%%%%%%%%%%%%%%%%%%%%%%%%%%%%%%%%%%%%%%%%%%%
\subsection{He $\rm I$ transmission spectra} \label{tran_spectra}
%%%%%%%%%%%%%%%%%%%%%%%%%%%%%%%%%%%%%%%%%%%%%%%%%%%

Our aim is to concentrate on the mid-transit absorption, which provides information about the main structure of the thermospheric escaping gas, but not to explain all the \het\ absorption features of those planets, nor their variations along the transit. 
In particular, the analysis of the velocity shifts in the \het\ absorption is potentially interesting because it would provide information about the 3D velocity distribution, its origin (e.g. day-to-night and super-rotation winds, and planet's rotation), and it may also break some of the degeneracy between the temperature and mass-loss rates. However, as our model is spherically symmetric, it cannot explain net blue or red shifts and, hence, such analysis is beyond the scope of this paper. Nevertheless, 
in order to obtain the best fit to the mid-transit spectra, we need to assume some net velocities 
along the observational LOS superimposed on the gas radial velocities of our model.
Thus, before exploring the range of temperature and mass-loss rates (see Sect.\,\ref{sec:chi2}), we first discuss the shape of the mid-transit spectra and the required additional velocities.

%%%%%%%%%%%%%%%%%%%%%%%%%%%%%%%%%%%%%%%%%%%%%%%%%%%
\subsubsection{\het\ absorption of \hdu18} \label{ana_hd189}
%%%%%%%%%%%%%%%%%%%%%%%%%%%%%%%%%%%%%%%%%%%%%%%%%%%

According to the observations by \cite{Salz2018}, \hdu18 shows a significant He excess absorption peaking at mid-transit (see Fig.\,\ref{absorption}), which is much stronger than that of \hd20 \citep[see][]{Alonso2019,Lampon2020}. 
The absorption profile also exhibits a more pronounced displacement to bluer wavelengths ($-3.5\pm$0.4\,\kms) and it is also significantly broader.
A similar net blue shift has been observed in the \het\ absorption of
\gj34\ \cite[see below and][]{Palle2020}. Then, in order to obtain the best possible fit, we incorporated a net blue shift of $-$3.5\,\kms\ in our simulations.

A further analysis of the spectrum shows that, at typical thermospheric temperatures for this planet, $\sim$12\,000\,K 
\cite[see, e.g.][]{Guo_2011,Salz2016,Odert_2020}, the Doppler broadening is insufficient to explain the width of the absorption profile (see Model A in left panel of Fig.\,\ref{absorption}). To achieve a similar width, we would need temperatures much higher than 20\,000\,K, which do not seem very realistic. 
Including a turbulence broadening component (see Sect.\,\ref{modelling_absorption}), the profile broadens, but it is still narrower than the measured absorption.  
In that calculation we also included the component of the gas radial velocities of our model along the observation LOS. However, since the absorption is confined to the first few thousands of kilometres and our velocities at these altitudes are rather slow (see left panels of Figs.\,\ref{he3} and \ref{vel}), the induced broadening is negligible. Hence, our hydrodynamic model alone is not able to explain the width of the absorption profile.

A likely explanation of the broadening emerges from the inspection of the observed shifts of the absorption profile during the ingress and egress transit phases \citep[Sect.\,\ref{observations},][]{Salz2018}. 
These shifts can be produced by a combination of the planetary rotation and strong net winds (probably of day-to-night and super-rotational winds) 
\citep[see e.g.][]{Salz2018,Flowers_2019,Seidel2020},
at altitudes $\sim$1--2\,\rp, where the \het\ absorption mainly takes place (see left panel of Fig.\ref{he3}).
In particular, \cite{Salz2018} derived an averaged wind in the range of $-$11.6 to $-$13.6\,\kms\ from the blue shift in the egress, and an averaged wind in the range of 3.4 to 9.6\,\kms\ from the red shift in the ingress.
Thus, we fitted the absorption by including, in addition to the main atmospheric blue component at $-$3.5\,\kms, a blue and a red atmospheric component. 
By perturbing the velocity and fractional contribution of those components and minimising the $\chi^2$, we found that the absorption profile can be well reproduced by two components at $-$12\,\kms\ and 5\,\kms, covering about 25\% and 28\% of the disk, respectively (Models B and C in the left panel of Fig.\ref{absorption}). We note that those velocities are very similar to the mean values of the winds derived by \cite{Salz2018} from the egress and ingress phases. 
Also our red-shifted atmospheric fraction agrees very well, although the blue-shifted fraction is smaller in our case (mid-transit) than in the egress.
These three components are included when we analyse the best fit to the spectra, for example in Model C (see Sect.\,\ref{sec:chi2}).

Our model can also fit the absorption in the weaker \het\ line (near 10832\,\AA) 
(see Model C in left panel of Fig.\,\ref{absorption}). This is due to the assumption of  a lower boundary with a  high enough density so that it absorbs all the XUV flux reaching this altitude (compare Models C, with a boundary condition, and B, without a boundary condition, in that figure). 
As the strong stellar radiation reaches the lower boundary of the upper atmosphere of \hdu18 (see left panel of Fig.\,\ref{hvmr_candidates}), and since the stronger lines are saturated at low altitudes, the relative absorption of the stronger lines with respect to the weaker lines decreases. 

However, in order to fit the \he_rat ratio, we had to increase the lower boundary density up to 10$^{18}$\,cm$^{-3}$, which agrees with the result of a very compressed annulus suggested by \cite{Salz2018}. 
At this high concentration, collision processes become more important, increasing the production 
of \het\ via recombination, because of the increase in He$^+$ by charge-exchange, 
and by electron collision \cite[see Table 2 in][]{Lampon2020}. These processes are more important at high temperatures (see inset in Fig.\,\ref{he3}). 
Overall, the fitting of the \he_rat\ ratio in this planet requires a 
very compressed and hot lower boundary.

It is worth noting that our model reproduces the measured \he_rat ratio of 2.8 better than the simple annulus model of \cite{Salz2018}, who obtained a value of 4.6. The key difference is that while the annulus is optically thick to the stronger lines in all its extension, 1.2\,\rp, they are only optically thick at the lower altitudes in our model.

%%%%%%%%%%%%%%%%%%%%%%%%%%%%%%%%%%%%%%%%%%
\subsubsection{Comparison with previous estimations of gas radial velocities of \hdu18} \label{seidel}
%%%%%%%%%%%%%%%%%%%%%%%%%%%%%%%%%%%%%%%%%%

\cite{Seidel2020} have recently re-analysed observations of Na in \hdu18 and have retrieved vertical upward winds, for example, radial velocity winds, of 40$\pm$4\,\kms\ at altitudes above 1\,$\mu$bar. The region probed by them, however, is limited to below $\sim$16\,000\,km (see their Fig.\,8) (1.2\,\rp\ referred to the centre of the planet), that is, the 1\,$\mu$bar to $\sim$1\,nbar or, approximately, (5--16)$\times 10^{3}$\,km or (1.06--1.2)\,\rp\ region. They also suggest that such high radial velocities could arise from the expanding thermosphere. Our model velocities are much lower than 40\,\kms\  (see Fig.\,\ref{vel}, left panel); although, they could be affected by the imposed null velocity at the base of our model at 1.02\,\rp. 

In order to verify if the shape of the measured \het\ absorption profile is compatible with the value derived by \citeauthor{Seidel2020}, we simulated an absorption profile with the \het\ densities derived in this work, but by assuming that the atmosphere is escaping at a constant radial velocity of 40\,\kms\ at all altitudes. 
The \het\ abundances that fit the absorption (see left panel of Fig.\,\ref{he3}) have a very pronounced peak at the lower altitudes of our model, 1--1.5\,\rp. Hence, in essence, that is equivalent to imposing a constant 40\,\kms\ velocity in that region. In order to be conservative, we did not include the turbulent component of the Doppler broadening in this calculation. The results show (see Fig.\,\ref{absorption}, Model D in green) that at such a high velocity, the \het\ absorption profile would be much broader than measured. At those altitudes, (1.06--1.2)\,\rp, the atmosphere is still dense enough to drag all atoms at a similar radial velocity \citep{Murray-Clay2009}. That is to say the velocities derived from either He or Na should not differ significantly. Hence, we conclude that such a high radial (vertical) velocity of 40$\pm$4\,\kms\ is not compatible with the \het\ measurements, but this rather suggests that the gas radial velocities at those altitudes are significantly smaller.

%%%%%%%%%%%%%%%%%%%%%%%%%%%%%%%%%%%%%%%%%%%%%%%%%%%
\subsubsection{\het\ absorption of \gj34} \label{ana_gj34}
%%%%%%%%%%%%%%%%%%%%%%%%%%%%%%%%%%%%%%%%%%%%%%%%%%%
\gj34\ shows a significantly larger He excess absorption than \hdu18 (we note the different y-axis scale in Fig.\ref{absorption}). Also, the absorption profile of the two stronger lines that are combined near 10833\,\AA\ is broader. As for \hdu18, we have included the Doppler and turbulence broadening  in the calculation. 
This warm Neptune has a weaker gravity that leads to a much more extended atmosphere which expands at larger velocities (see right panels of Figs.\,\ref{he3} and \ref{vel}). In fact, the velocities are already significant at rather low radii. 
Thus, in contrast to \hdu18, the component of the gas radial velocity along the observer LOS produces a significant broadening. We observe in Fig.\ref{absorption} (right panel) that the models that include the radial velocities (F and G), in contrast to E, explain very well the observed broadening without the need of blue or red components. 
We should also note that (not shown in the figure) the turbulent broadening is negligible compared to the broadening caused by such high radial velocities, and hence it does not have any significant impact on the line width.

As found in \hdu18\ (see above), as well as in \hd20\ \citep{Alonso2019}, we also observe a significant blue shift of the whole absorption, estimated to be $-$3.2$\pm$1.3\,\kms\ \citep{Palle2020}. 
This is intermediate between the values measured in those two planets. Unfortunately, we do not have ingress or egress spectra with a sufficient S/N (as we did for \hdu18) 
to help us in understanding its origin.
A likely explanation is that it is also produced by day-to-night winds with velocities along the LOS of $-$3.2\,\kms\ (see model~F in Fig.\,\ref{absorption}). However, as \gj34\ has a very extended atmosphere, with significant \het\, absorption beyond the Roche lobe (see left panel of Fig.\ref{he3}), another plausible 
interpretation is an upper atmosphere with no significant day-to-night winds below the Roche radius 
(3.6\,\rp, see Table\,\ref{eqw}) but with the unbound gas above the Roche lobe blue-shifted by processes, such as stellar wind interactions or stellar radiation pressure \cite[see e.g.][]{Salz2016}. Model~G (red curve in right panel of Fig.\,\ref{absorption}) shows the absorption of those simulations assuming that the gas above the Roche lobe is escaping at an LOS velocity of $-$5.0\,\kms. This scenario is consistent with  
that proposed by \cite{Bourrier2018} to explain the blue-shifted absorption signature of their \lya\ observations.

The absorption of the weaker \het\ line near 10832\,\AA\ is also significantly broadened, and it is well reproduced within the estimated error bars. We note though that, in contrast to \hdu18, the stronger lines are not saturated at any radii and hence the ratio of the strong to the weaker lines is larger in this planet. Thus, the \het\ absorption of this planet is rather insensitive to its lower boundary atmospheric conditions. 

\begin{figure*}
\includegraphics[angle=0.0, width=1.\columnwidth]{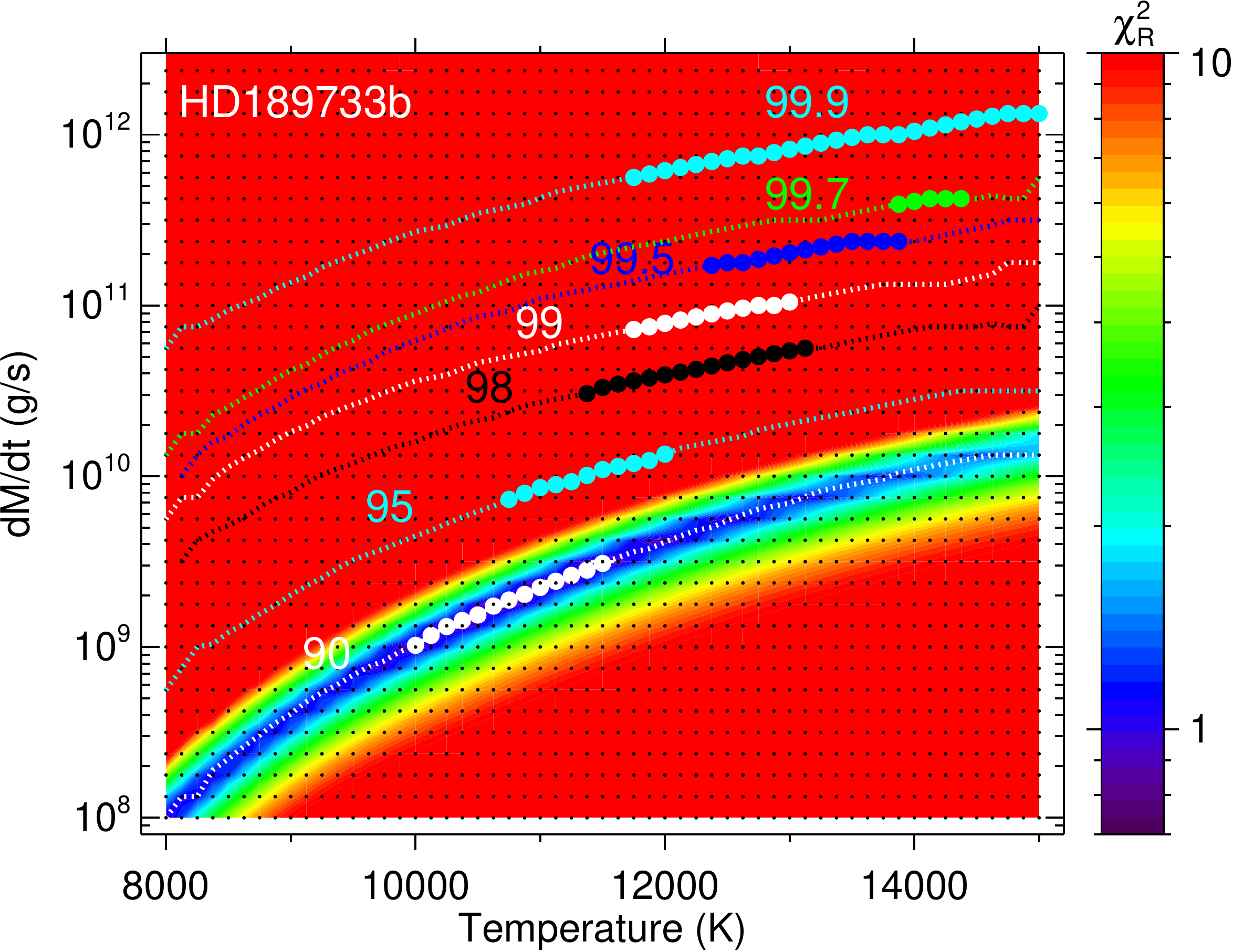}
\includegraphics[angle=0.0, width=1.\columnwidth]{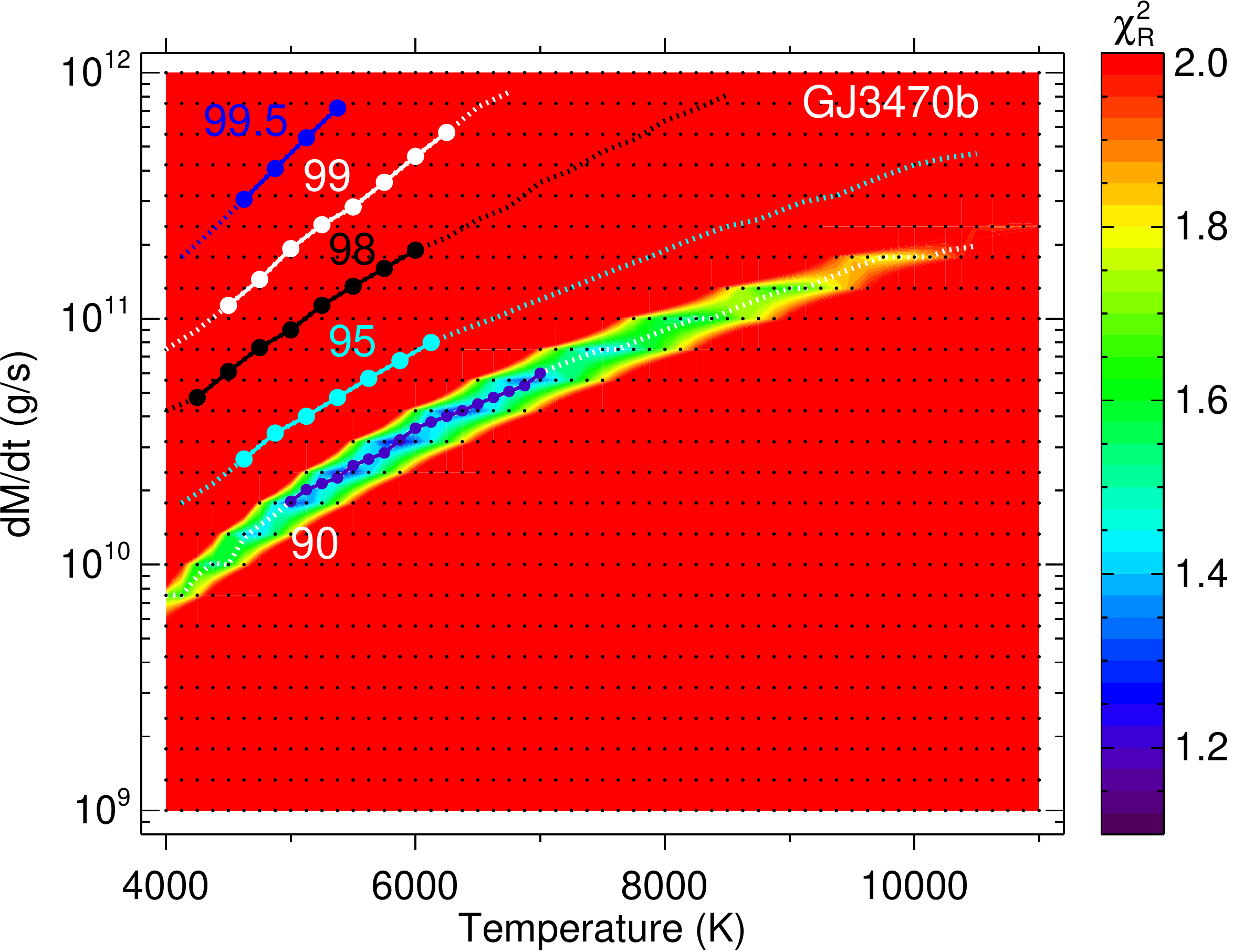} 
\caption{Contour maps of the reduced $\chi^2$ of the \het\ absorption for \hdu18 (left panel) and \gj34\ (right panel) for an H/He ratio of 90/10.
We note the different scales of temperatures and \mlr. Dotted curves represent the best fits with filled circles denoting the constrained ranges for a confidence level of 95\% (see Sect.\,\ref{model_grid}). 
Over-plotted are also the curves and symbols for several H/He ratios, as labelled. The labels correspond to the hydrogen percentage, e.g. `90' for an H/He of  90/10 and `95' for  H/He=95/5 (see  Sect.\,\ref{H_He_constraints}). The black dots represent the grid of the simulations.} 
\label{chi2}
\end{figure*}

\begin{figure}
\includegraphics[angle=0.0, width=1\columnwidth]{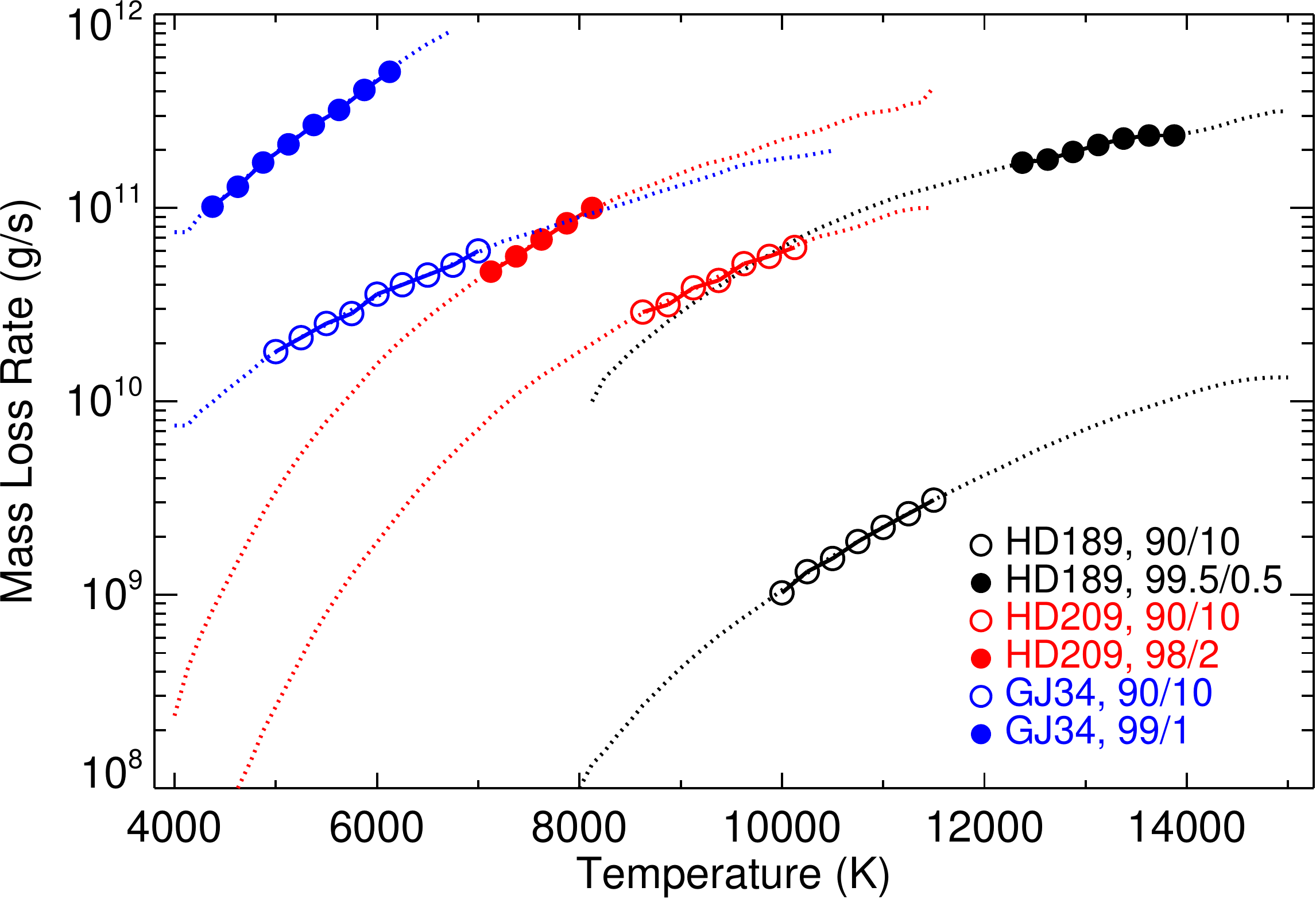} 
\caption{Ranges of temperature and mass-loss rates for \hd20, \hdu18, and \gj34 for H/He ratios of 90/10 and 99.5/0.5, 90/10 and 98/2, as well as 90/10 and 99/1, respectively.
Dotted lines show the ranges explored and symbols correspond to the constrained ranges (see Sect.\,\ref{model_grid}). 
The values for \hd20 were taken from \cite{Lampon2020}. The limited ranges (symbols) for this planet were obtained from heating efficiency considerations.}  
\label{mesos_chi2}
\end{figure}

%%%%%%%%%%%%%%%%%%%%%%%%%%%%%%%%
\subsection{Constraining the temperatures and mass-loss rates by the $\chi^2$ analysis} \label{sec:chi2}
%%%%%%%%%%%%%%%%%%%%%%%%%%%%%%%%

The $T$-\mlr\ curves dictated by the \het\ absorption of both planets show the typical behaviour of a positive correlation
(see Fig.\,\ref{chi2}).
That is to say for a given \het\ concentration (imposed by the measured absorption profile), if temperature increases, the \het\ concentration decreases and its maximum tends to move to lower altitudes (with smaller effective absorption areas) which, in order to be balanced, requires an increase in \mlr.

Different H/He compositions of the thermospheric gas show different $T$--\mlr\ curves, as was studied 
for \hd20 by \cite{Lampon2020}. For a given mass-loss rate and temperature, the effect of increasing the H/He ratio results in a decrease in the global mass density, and the hydrogen and \het\ concentrations, \citep[see Fig. 13 of][]{Lampon2020}. Then, to compensate for the \het\ absorption, the mass-loss rate has to be increased. In summary, for a fixed temperature, the higher the H/He ratio is, the higher the mass-loss rate required to reproduce the \het\, absorption.

\begin{figure*}[htbp]
\includegraphics[angle=0.0, width=1.0\columnwidth]{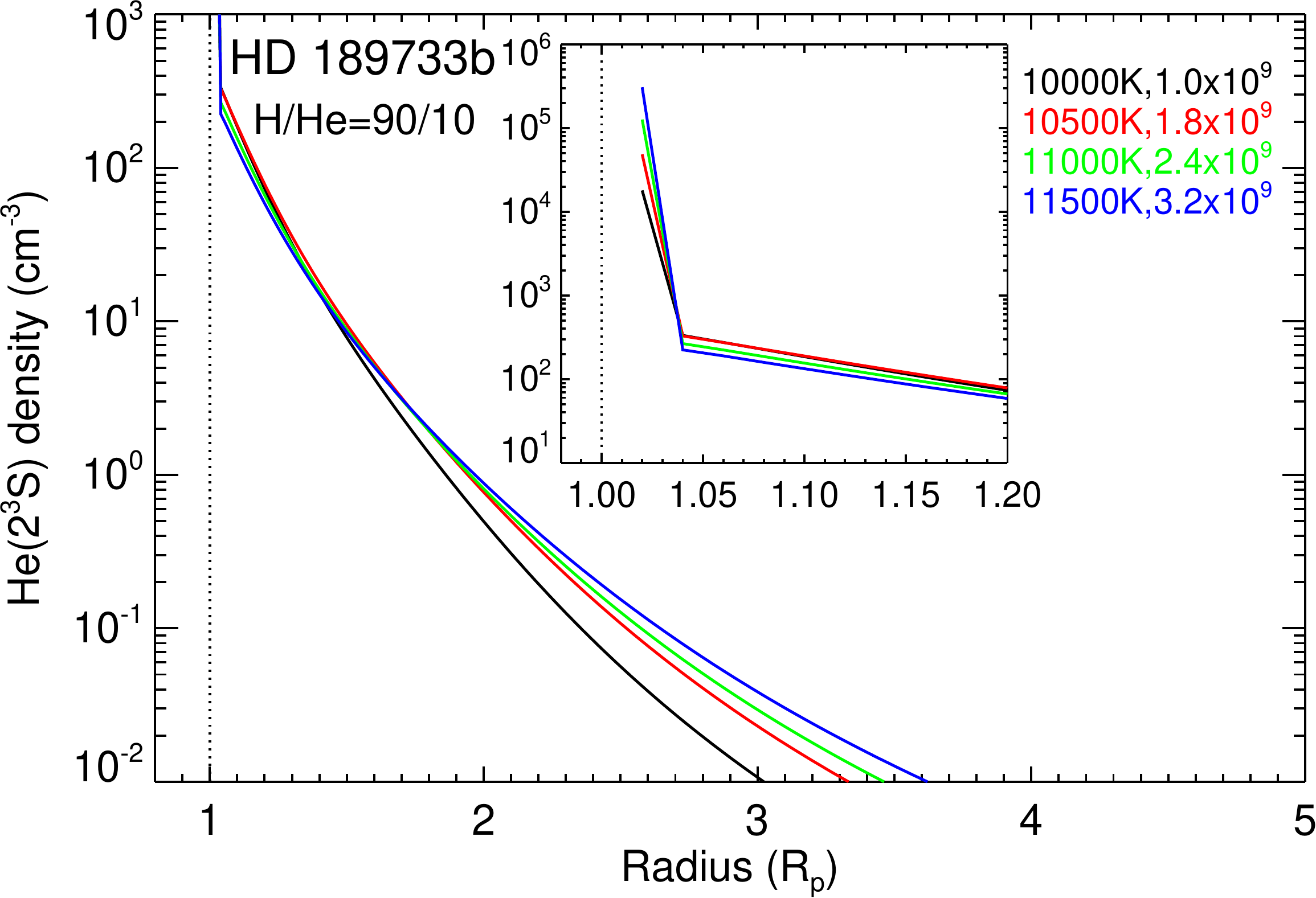}
\includegraphics[angle=0.0, width=1.0\columnwidth]{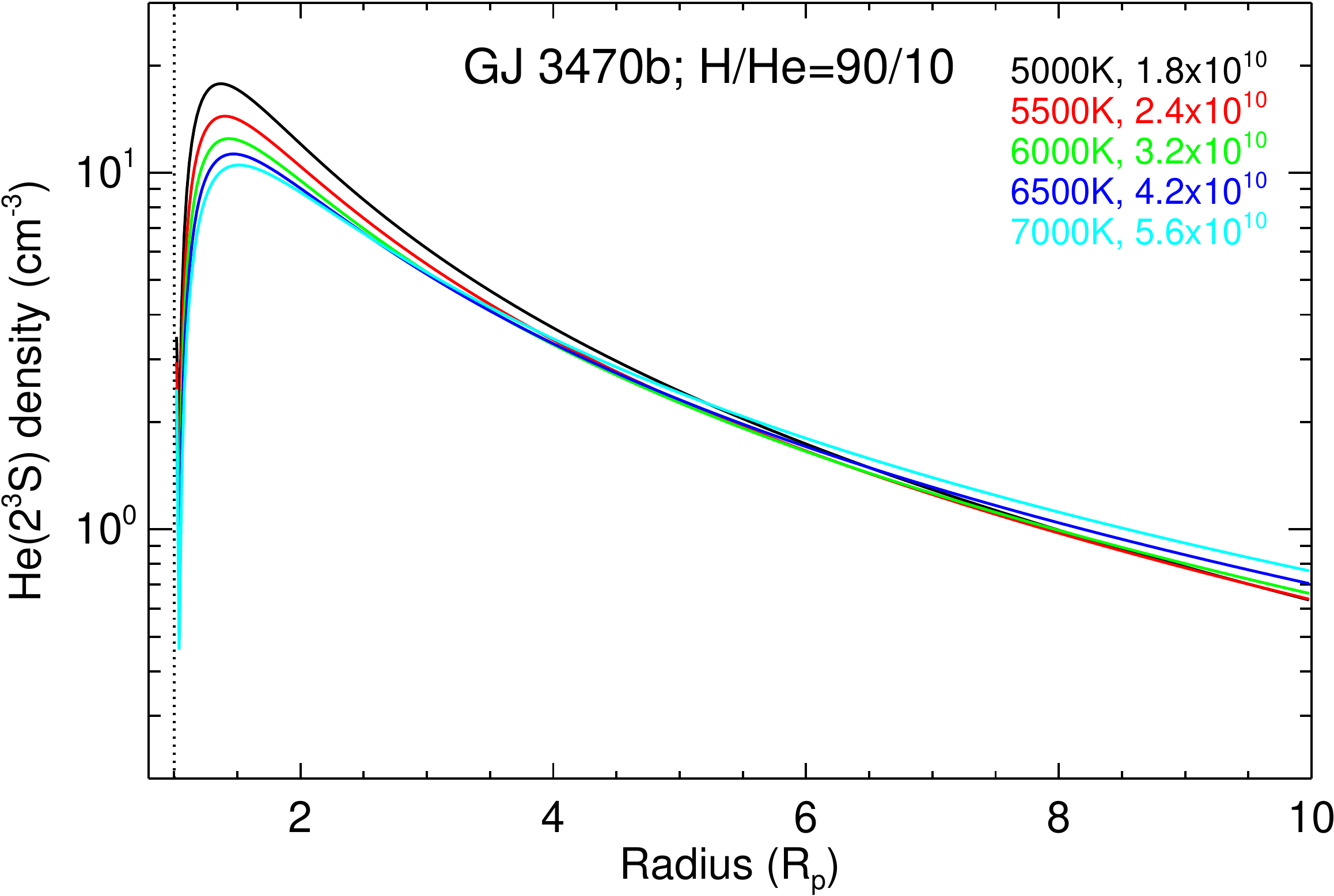}     
\caption{\het\ concentration profiles that best fit the measured absorption; i.e. for the white (\hdu18) and dark blue (\gj34) filled circles in Fig.~\ref{chi2}. We note the different scale of the x-axis. The inset in the left panel shows a zoom at low radii.} 
\label{he3} 
\end{figure*}

\begin{figure*}[htbp]
\includegraphics[angle=0.0, width=1.0\columnwidth]{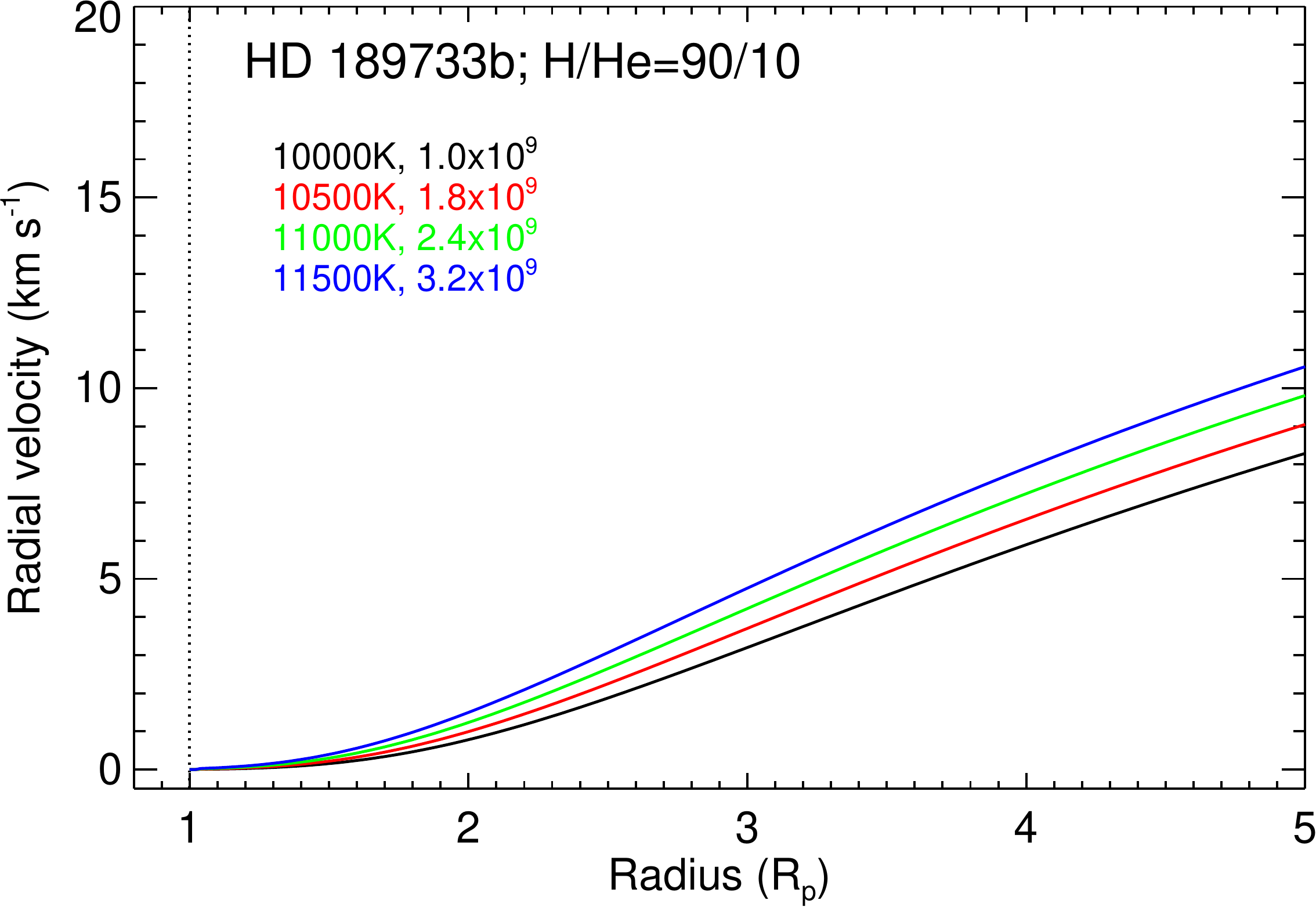}
\includegraphics[angle=0.0, width=1.0\columnwidth]{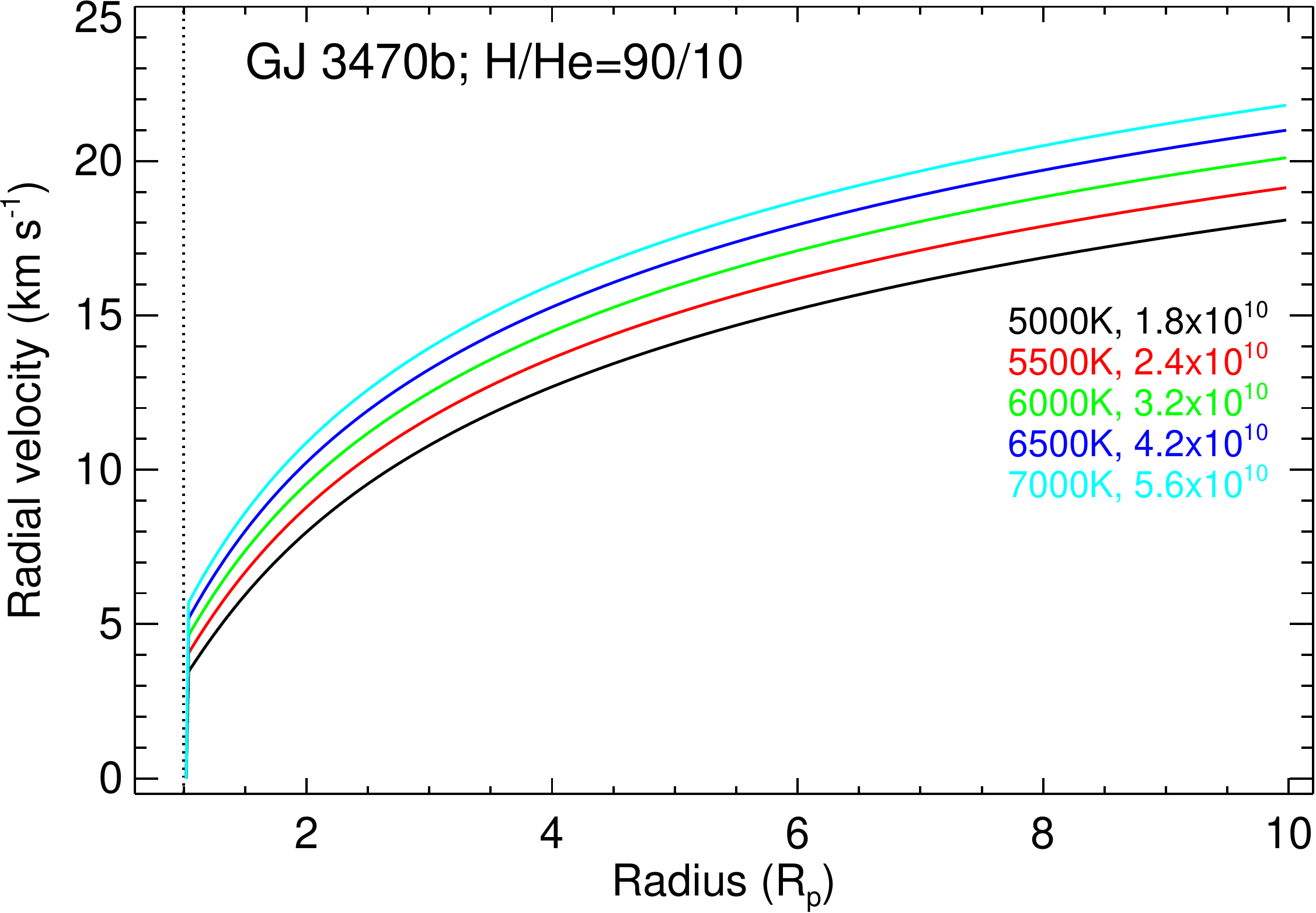}
\caption{Gas radial velocities of the model for the 
best fits of the \het\ measured absorption; i.e. for the white (\hdu18) and dark blue (\gj34) filled circles in Fig.~\ref{chi2}. We note the different scale of the x-axis.} 
\label{vel} 
\end{figure*}

\begin{table}
\centering
\caption{\label{eqw}Planet parameters, the XUV flux, and EW(\het) absorption.}
\begin{tabular}{l c c c } 
\hline  \hline  \noalign{\smallskip}
Planet                  & \hd20 & \hdu18  & \gj34 \\
\hline \noalign{\smallskip}
Mass ($M_{\rm Jup}$)    & 0.685\,\,$^{+0.015}_{-0.014}$ & 1.162\,\,$^{+0.058}_{-0.039}$ & 0.036\,\,$^{+0.002}_{-0.002}$ \\
\noalign{\smallskip}
Radius ($R_{\rm Jup}$)  & 1.359\,\,$^{+0.016}_{-0.019}$ & 1.23\,\,$^{+0.03}_{-0.03}$ &
0.36\,\,$^{+0.01}_{-0.01}$\\
\noalign{\smallskip}
Gravity ($g_{\rm Jup}$) & 0.371 &  0.768 &  0.278\\
\noalign{\smallskip}
$\Phi$$^{a}$  ($\Phi_{\rm Jup}$)        & 0.504 & 0.944 &  0.100\\
\noalign{\smallskip}
$R_{\textrm{lobe}}$ (\rp)$^{b}$ & 4.2 &  3.0 &  3.6\\
\noalign{\smallskip}
$F_{\rm XUV}$$^{c}$ & 2.358     &  56.74                &  3.928        \\
\noalign{\smallskip}
EW (m\AA)$^{d}$ & 5.3$\pm$0.5   &  12.7$\pm$0.4 & 20.7$\pm$1.3  \\
\noalign{\smallskip}
\hline
\end{tabular}
\tablefoot{
Planetary mass and radius of \hd20 from \cite{Torres08}. Those of \hdu18 and \gj34 are taken from Table\,\ref{table.parameters} and included here for easier comparisons. 
\tablefoottext{a}{Gravitational potential.} 
\tablefoottext{b}{Roche lobe of \hd20, \hdu18, and \gj34 by \cite{Salz2016}, \cite{Eggleton_1983}, and \cite{Bourrier2018}, respectively. }
\tablefoottext{c}{XUV flux in units of 10$^{3}$ erg\,cm$^{-2}$\,s$^{-1}$ at $\lambda$ < 912 \AA\ at planetary distance, calculated from \cite{Lampon2020} for \hd20 and from Fig.\,\ref{flux_xsec} for \hdu18 and \gj34.} 
\tablefoottext{d}{Equivalent width (EW) integrated in the range of 10831.0$-$10834.5\,\AA.} 
}
\end{table}

In comparing the results of \hdu18 and \gj34 (Fig.\,\ref{chi2})  with those of \hd20  \citep[Fig.~8 of][]{Lampon2020}, we can appreciate that the $T$-\mlr\ curves of \hdu18\ and \gj34\ are better constrained.
In the case of \hdu18, the reduction of the degeneracy comes from the fitting of the \he_rat\ ratio and from the temperature broadening, principally when including the turbulence.

The minima of $\chi^2_R$ for \hdu18 are larger at temperatures below $\sim$10\,000\,K and above $\sim$12\,500\,K. 
For lower temperatures, despite the high density in the lower boundary of this planet, \het\ density is too low for fitting the weaker line, similar to Model B in Fig.\,\ref{absorption}; while at high temperatures, it is well fitted (see Sect.\,\ref{ana_hd189}). 
At temperatures above $\sim$12\,500\,K, the lines are too broadened, particularly when including the turbulence, and then the fitting is worse. It is interesting to look at the T/\mlr\ constrain when neglecting the turbulence (see Fig.\,\ref{chi2_noturb}). We see that the constrain changes significantly, leading in general to larger temperatures and mass-loss rates although the fit is not so good (larger minimum $\chi^2$).

As the strong line of \hd20\ and \gj34\ is not saturated at any altitude, the \he_rat\ ratio does not contribute to reduce the model degeneracy for these exoplanets. 
For \gj34, as the broadening of the gas radial velocities is very large, the effects of the turbulence are negligible.

In the case of \gj34, the reduction of the degeneracy comes from the large radial velocities (see right panel of Fig.\,\ref{vel}).  As we can see in Fig.\,\ref{chi2} (right panel), $\chi^2$ is worse at temperatures below about 5400\,K and higher than $\sim$6900\,K. At lower temperatures,  the velocities are smaller and the broadening of the absorption profile narrower. The opposite occurs for higher temperatures. Thus, radial velocities calculated by the hydrodynamic model help to constrain the $T$-\mlr\ curves of \gj34.
However, \hd20 and \hdu18 do not have such high radial velocities, so they do not help to reduce the degeneracy in these exoplanets.

We note that as the $T$-\mlr\ degeneracy of \hd20\ is not reduced by fitting the \he_rat\ ratio nor by the gas radial velocities,
\cite{Lampon2020} reduced it by applying constraints on the heating efficiencies. 
This criterion, however, cannot be used to reduce the degeneracy in \hdu18 as heating efficiencies in this exoplanet are rather uncertain because of the significant radiative cooling
\citep[see, e.g.][]{Salz_2015,Salz2016}.

We have found that the mass-loss rates of \gj34 are much larger than those of \hdu18 (Fig.\,\ref{chi2}); they are more than a factor of 10 for similar temperatures and an H/He ratio of 90/10. For larger H/He ratios, this difference decreases. Using the  $T$-\mlr\ curve of \hd20 as a reference \citep[reproduced in Fig.\,\ref{mesos_chi2}]{Lampon2020}, we observe that the  corresponding curves for \hdu18 and \gj34 are located in opposite regions, below and above that of \hd20. On the one hand, the mass-loss rate of \hdu18 is more than one order of magnitude smaller than that of \hd20, and it is located at higher temperatures. The larger XUV flux of \hdu18 (see Table\,\ref{eqw}) favours its larger temperatures. However, its larger gravity prevents its evaporation, resulting in much lower mass-loss rates.
On the other hand, \gj34 is not only irradiated in the XUV at higher levels than \hd20 (although only slightly), but it also has a much smaller gravity (see Table\,\ref{eqw}). Both factors favour its larger mass-loss rate, although the second is by far the most important. Those factors lead to a very extended atmosphere for \gj34 and a rather compressed one for \hdu18 (see Fig.\,\ref{he3}); while the \het\ concentration drops by a factor of 10 in $\sim$7\,\rp\ for \gj34, it takes only $\sim$0.5\,\rp\ for \hdu18. Likewise, the velocities of atmospheric expansion of \gj34 are significantly larger than those of \hdu18 (see Fig.\,\ref{vel}). It is important to note that for \gj34, the velocities are already significant at very small radii, which, together with the large \het\ abundances at these distances, contribute to a significant broadening of the absorption profile (see Sect.\,\ref{ana_gj34}).

\begin{figure*}
\includegraphics[angle=0.0, width=1.0\columnwidth]{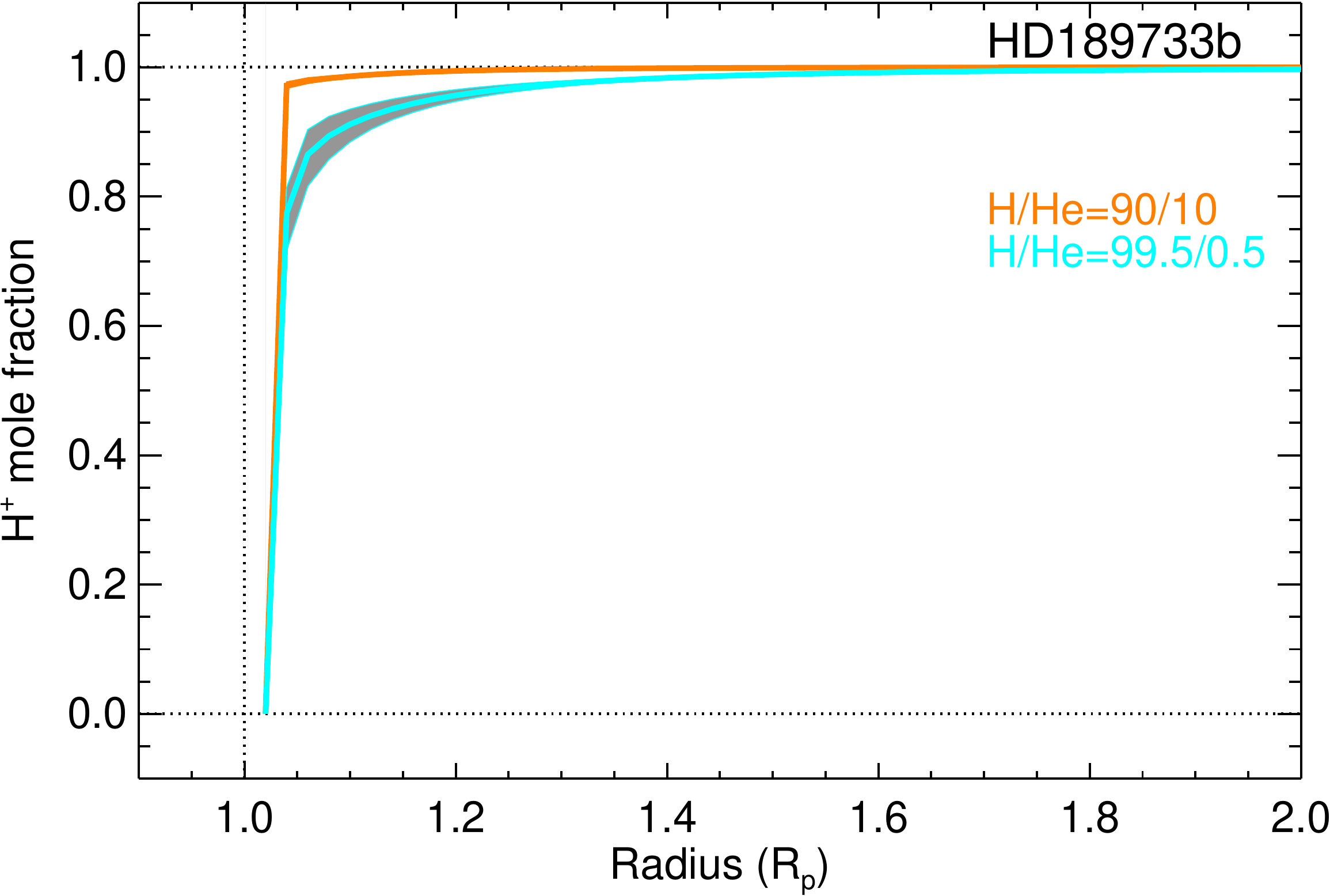}
\includegraphics[angle=0.0, width=1.0\columnwidth]{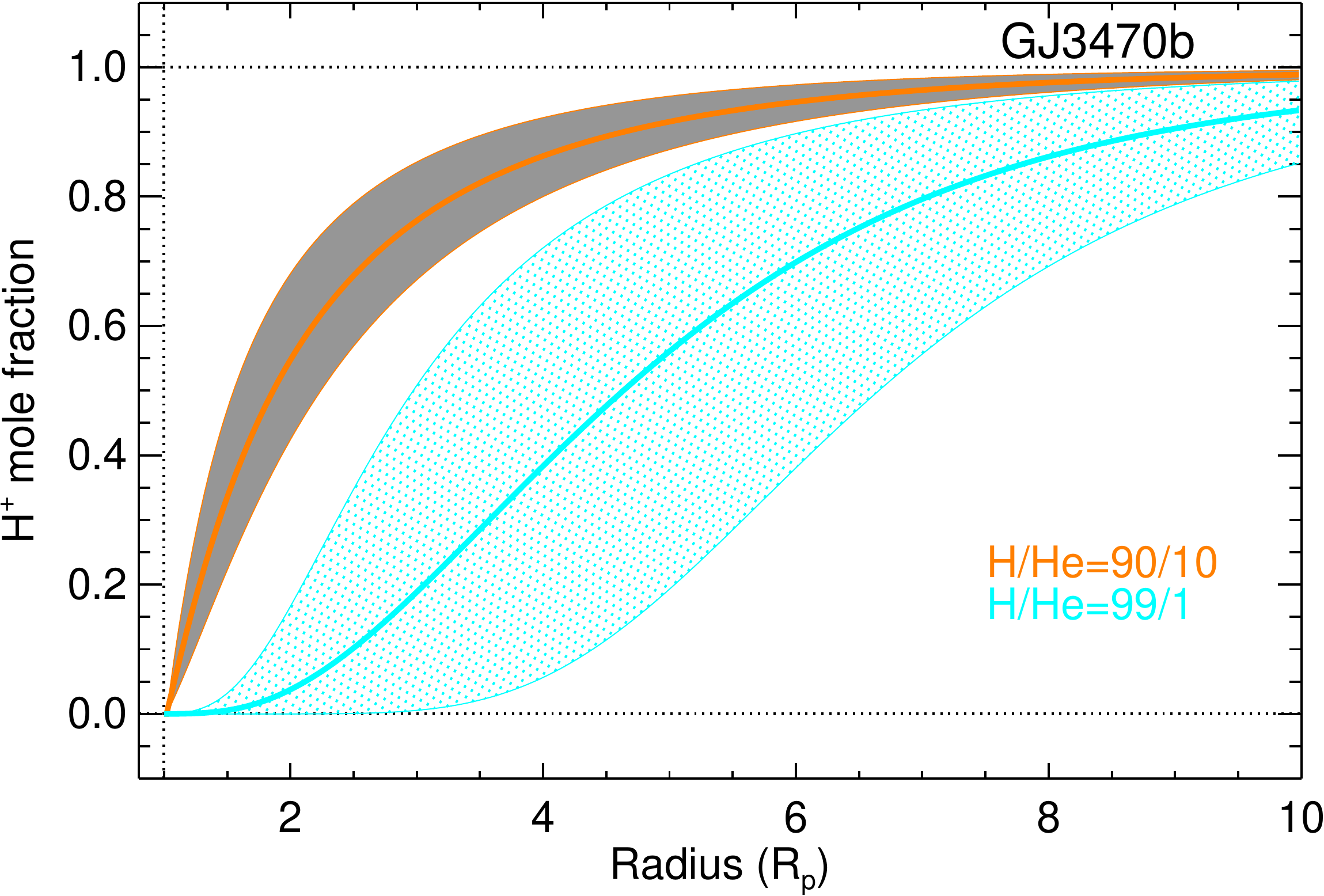}
\caption{H$^+$ mole fraction profiles resulting from the fit of the measured absorption (filled circles in Fig.~\ref{chi2}) for H/He=90/10 and 99.5/0.5 for \hdu18 (left panel) and 90/10 and 99/1 for \gj34 (right panel). We note the different x-axis range.  The solid thicker lines are the mean profiles.}
\label{hvmr_candidates} 
\end{figure*}

%%%%%%%%%%%%%%%%%%%%%%%%%%%%%%%%
\subsection{\het\ density profiles and ionisation of the upper atmosphere}
\label{other_constrains}
%%%%%%%%%%%%%%%%%%%%%%%%%%%%%%%%

In this section we show the derived \het\ densities, gas radial velocities, and the ionisation state of the upper atmospheres of  \hdu18 and \gj34. The \het\ density profiles for our nominal case of H/He=90/10 are shown in Fig.\,\ref{he3} (the profiles for the derived H/He ratios are shown in Fig.\,\ref{he3_supp}).
The densities of \hdu18 peak at its lower boundary.
\gj34 shows more extended \het\ density profiles with lower values at lower altitudes and larger values at higher altitudes. The peaks of the density profiles for this planet are also well confined to the range of 1.3--1.5\,\rp.

The results for the gas radial velocities are shown in Fig.\,\ref{vel} for the nominal H/He=90/10, and in Fig.\,\ref{vel_supp} for the derived H/He ratios. \gj34\ is already expanding at large velocities at rather low altitudes, which is supported by the rather wide observed absorption profile. Fig.\,\ref{absorption} shows (right panel, Model E, orange) that if these radial velocities are not included, the \het\ line profile would be significantly narrower.

The resulting ionisation fronts can be seen in the H$^{+}$ mole fractions plotted in Fig.\,\ref{hvmr_candidates} for the nominal and the derived H/He ratios. The ionisation front of \hdu18 is closer to the planet's lower boundary and narrower than that of \gj34. That is, for \hdu18, the stellar flux is strongly absorbed in a narrow altitude interval near the lower boundary, while in the case of \gj34, it is absorbed progressively in a wide range of altitudes at relatively large distances. Nonetheless, the effective absorption radius, R$_{XUV}$, the distance where the optical depth for the XUV radiation is unity  \citep{Watson_1981}, is 1.02\,\rp\ (i.e. the lower boundary) for \hdu18 and is in the range (1.02--1.12)\,\rp\ for \gj34. As the planetary XUV cross section varies with R$_{XUV}^2$, the stellar radiation energy absorbed by \gj34 increases due to the hydrodynamic atmospheric escape, while that of \hdu18 remains constant.

\begin{figure*}
\includegraphics[angle=0.0, width=0.666\columnwidth]{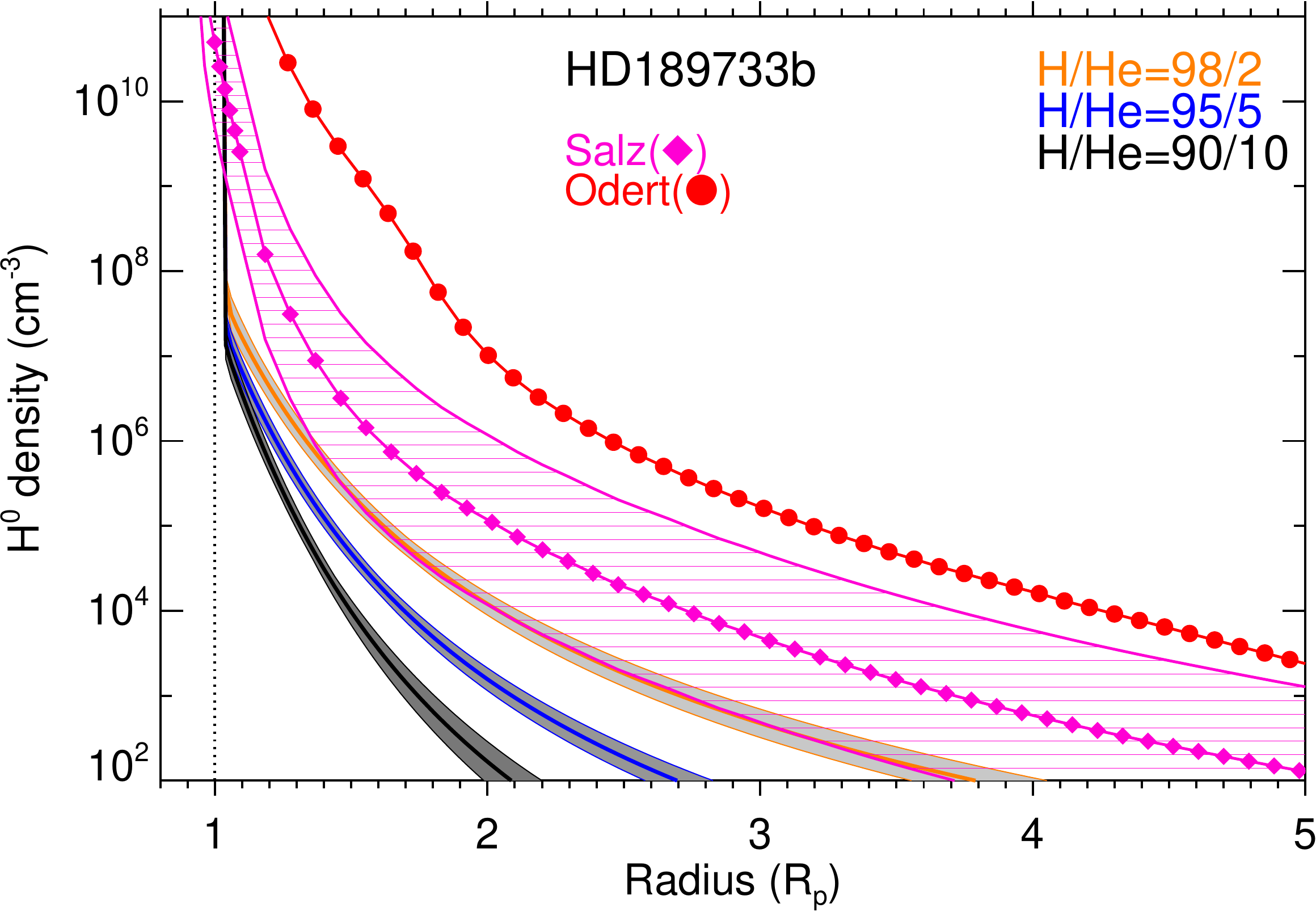}
\includegraphics[angle=0.0, width=0.666\columnwidth]{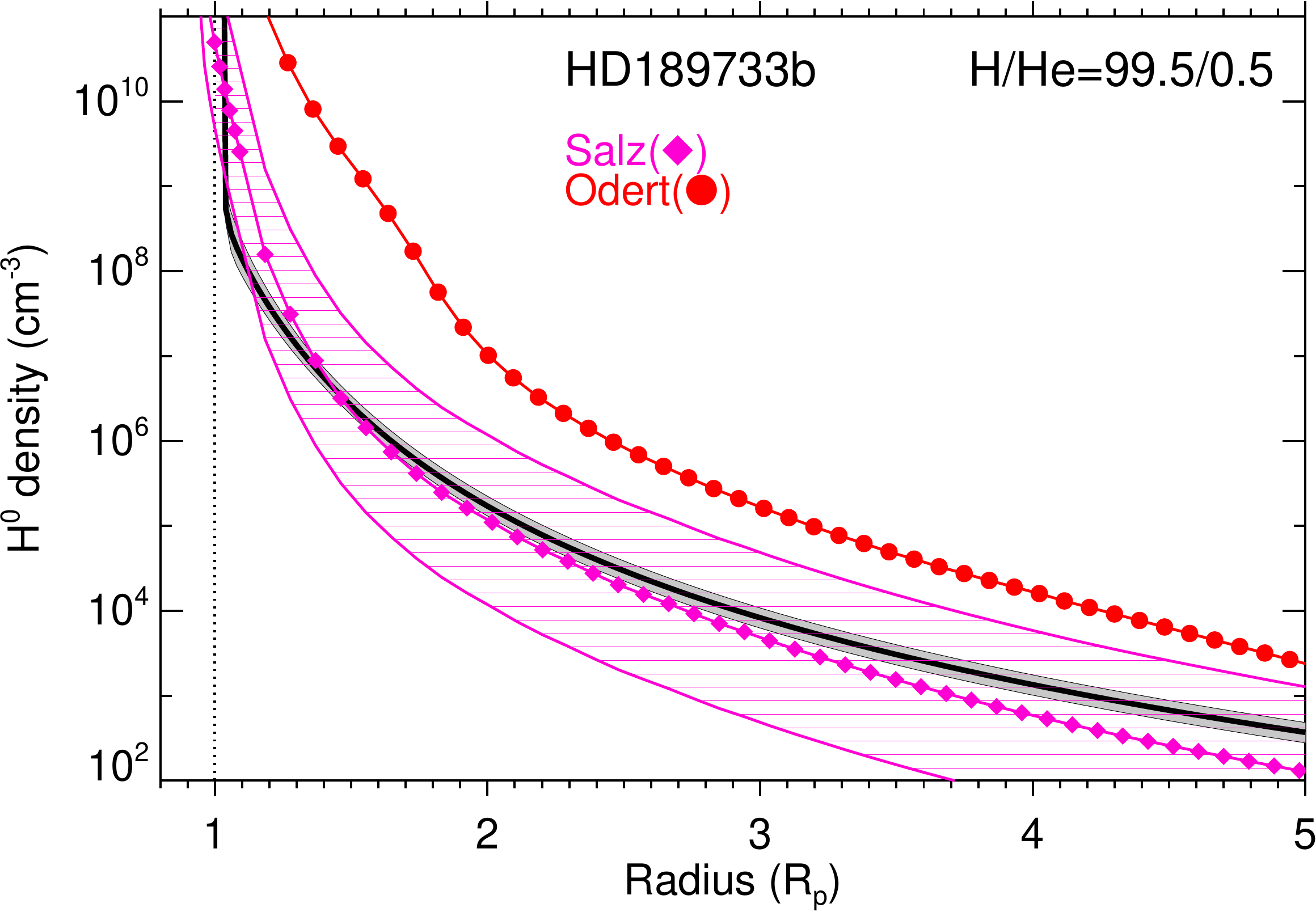}
\includegraphics[angle=0.0, width=0.666\columnwidth]{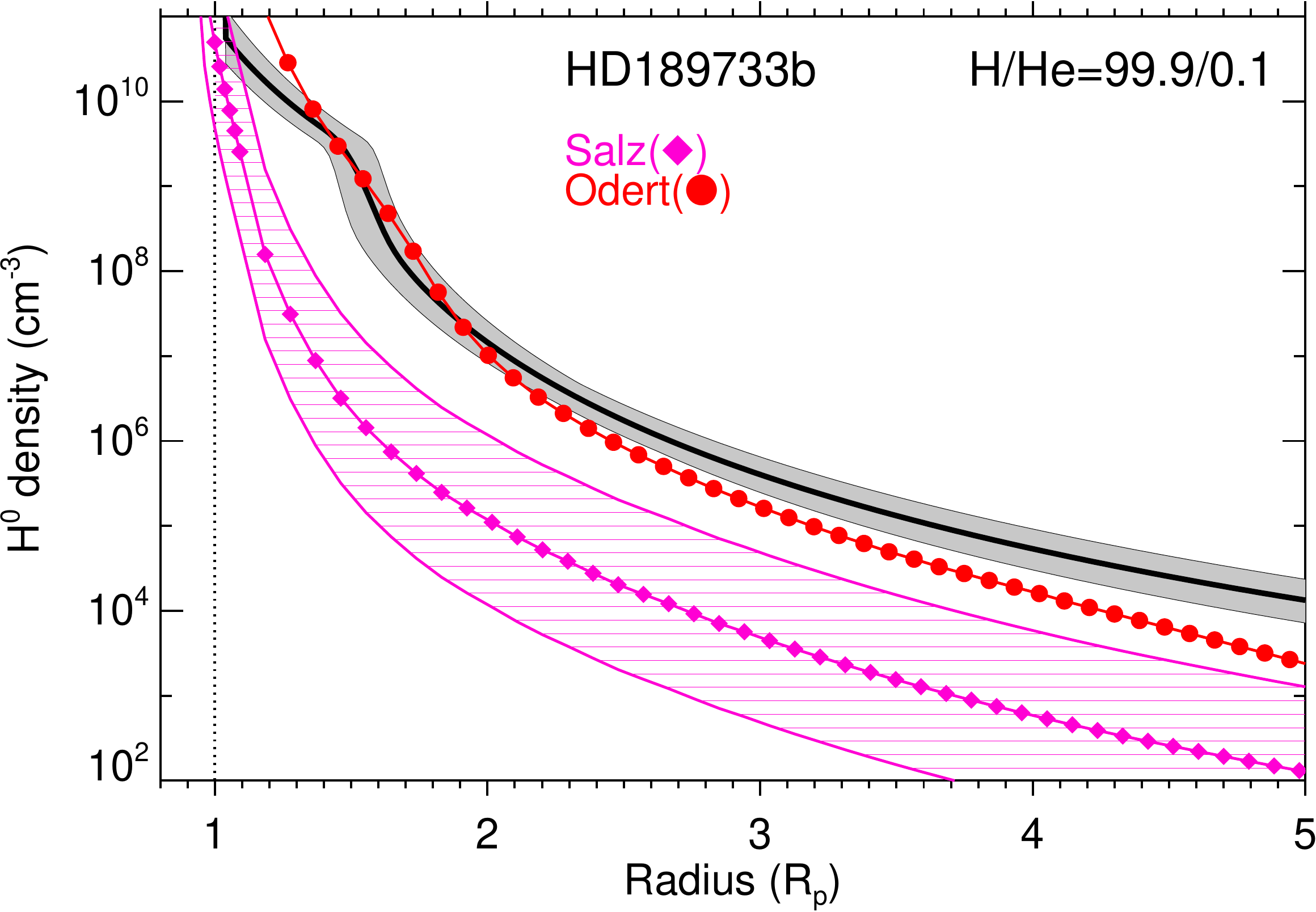}
\includegraphics[angle=0.0, width=0.68\columnwidth]{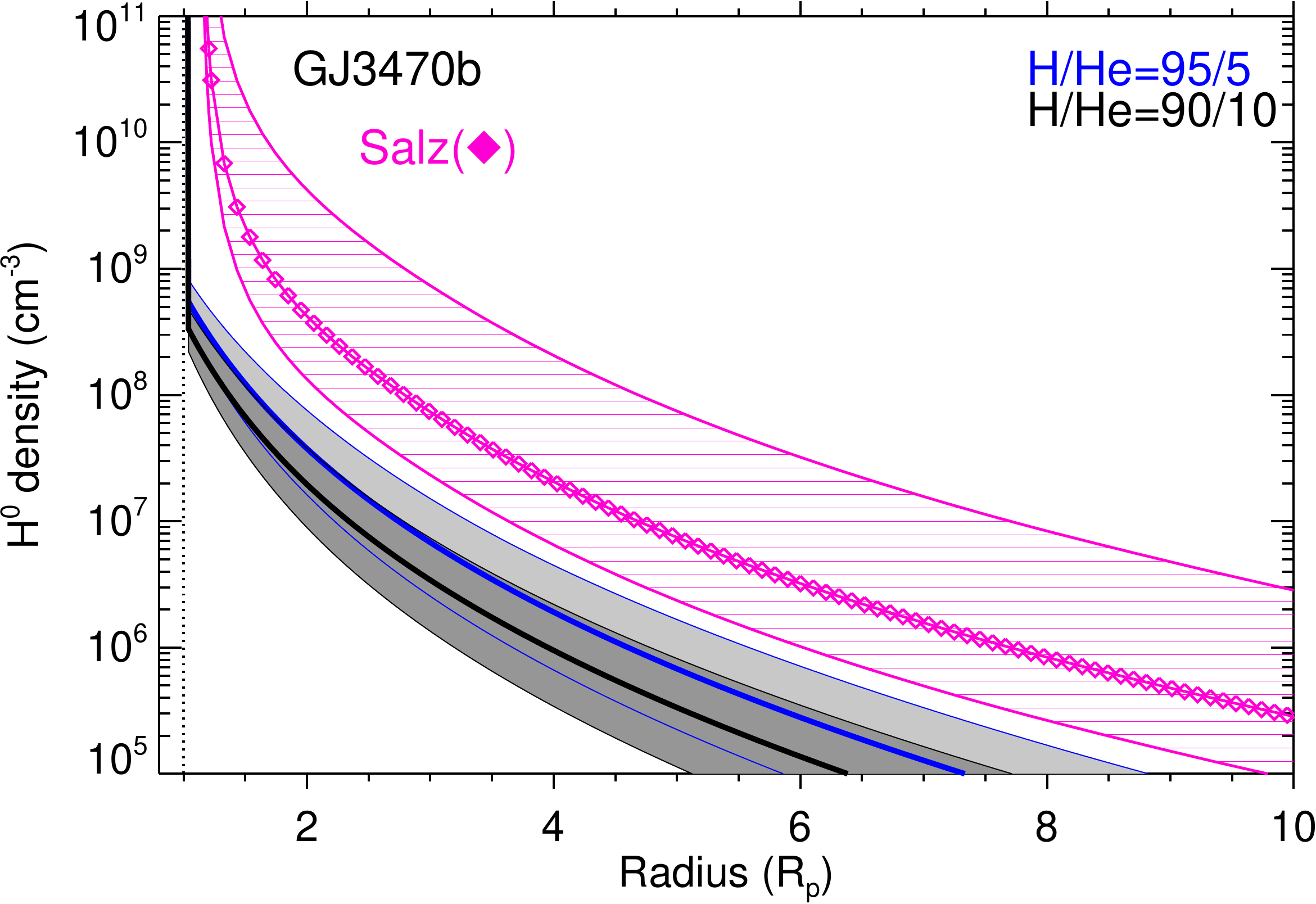}
\includegraphics[angle=0.0, width=0.68\columnwidth]{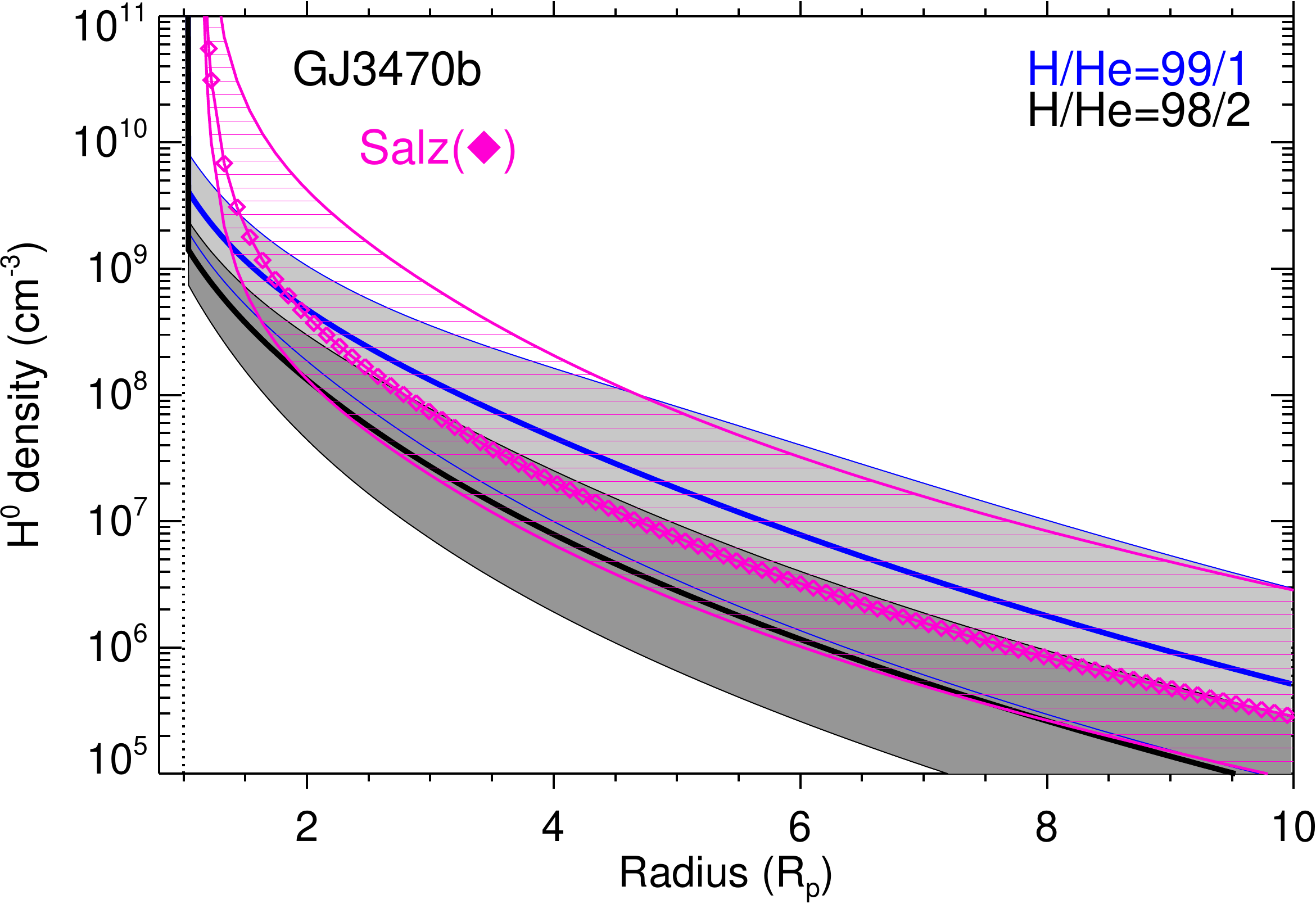}
\includegraphics[angle=0.0, width=0.68\columnwidth]{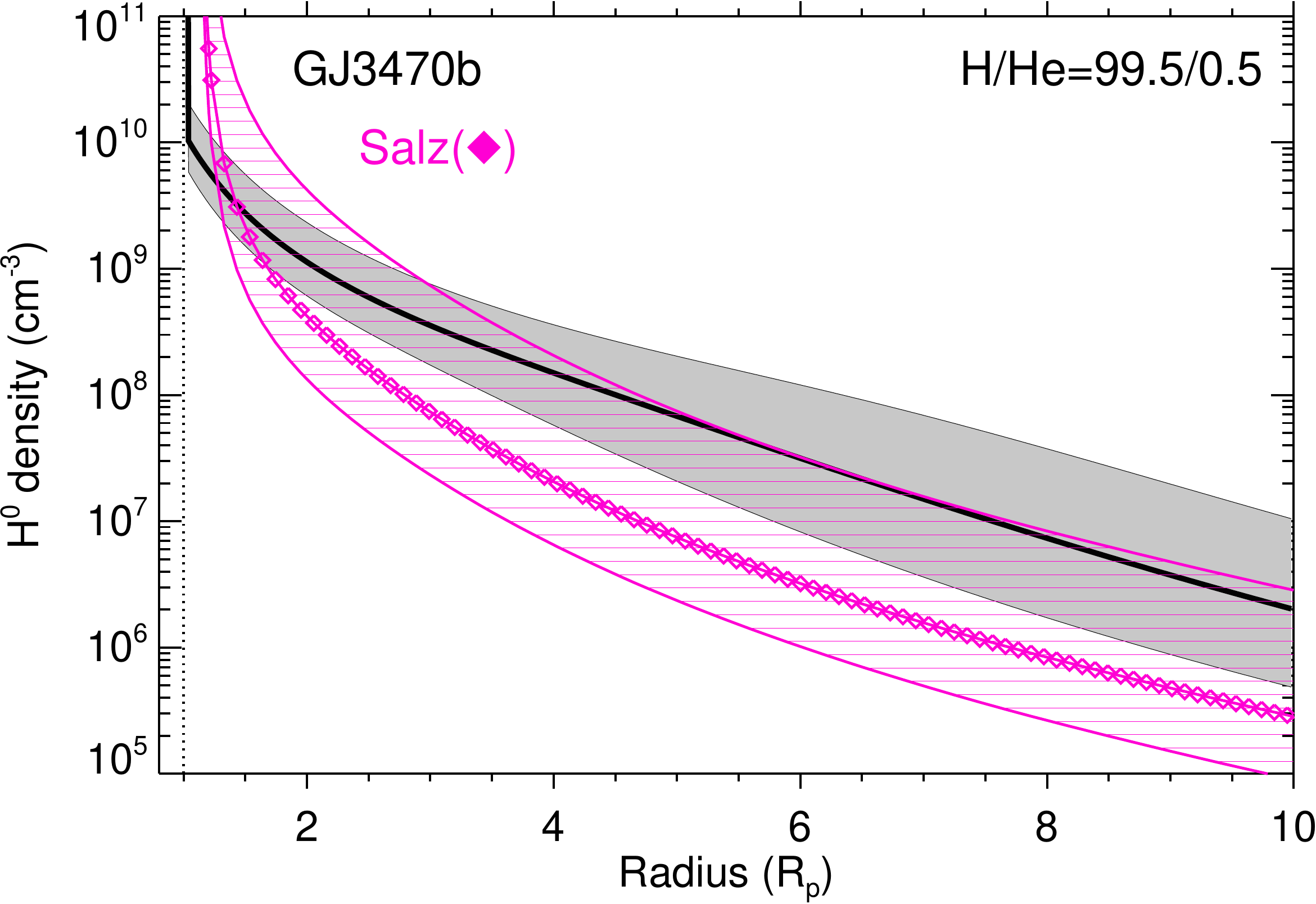}
\caption{ Range of the neutral hydrogen concentration profiles (in grey shade areas) resulting from the fit of the measured absorption (filled circles in Fig.~\ref{chi2})  
for \hdu18 (upper panels), \gj34 (lower panels), and several H/He ratios, as labelled. The solid thicker curves are the mean profiles. The H$^0$ density derived from \lya\ measurements for \hdu18 and \gj34 by \cite{Salz2016} and our estimated uncertainties ($\times$10 and /10 for \hdu18, as well as $\times$10 and /$\sqrt{10}$ for \gj34) are also shown (diamonds and hatched areas in magenta). The red profile is  the H$^0$ density derived by \cite{Odert_2020} from their SF11 model for \hdu18.}
\label{hden_candidates} 
\end{figure*}

%%%%%%%%%%%%%%%%%%%%%%%%%%%%%%%%
\subsection{Constraining the H/He composition by the $\chi^2$ analysis}
\label{H_He_constraints}
%%%%%%%%%%%%%%%%%%%%%%%%%%%%%%%%

As in \cite{Lampon2020}, we constrained the H/He ratio by matching the H$^0$ abundance profiles imposed by the \het\ observations with those derived from \lya\ absorption measurements.
By constraining the H/He ratio, we also reduced the $T$-\mlr\ degeneracy (see Fig.\,\ref{chi2}).

For \hdu18, we compared our H$^0$ abundance profiles to those retrieved by \cite{Salz2016} and \cite{Odert_2020} from \lya\ absorption measurements. 
These authors analysed the observations of \cite{Lecavelier_des_Etangs_2012} using a 1D hydrodynamic model and assuming substellar conditions to be representative of the whole planet.  
While \cite{Salz2016} report a good fit of the \lya\ absorption, \citeauthor{Odert_2020} slightly overestimate it. 
We performed \lya\ absorption calculations and found, in effect, that the profile of \cite{Salz2016} fit the observations of \cite{Lecavelier_des_Etangs_2012} better. In addition we obtained that the errors in these measurements are well embraced by the H$^0$ profile of \cite{Salz2016} when scaled by factors of 0.1 and 10 (see upper panels in Fig.\,\ref{hden_candidates}).

From the analysis of $\chi^2$ (i.e. the profiles that fit the measured \het\ within the minimum $\chi^2$ values, see Sect.\,\ref{model_grid}), we found that the H$^0$ model density profiles for H/He ratios of 90/10, 95/5, and even for 98/2 (see upper left panel of Fig.\,\ref{hden_candidates}) are significantly lower than that of \cite{Salz2016}, including their estimated uncertainties, at all altitudes. Also, for an H/He ratio of 99.9/0.1 and larger values, the H$^0$ \lya\ density profile is clearly overestimated (upper right panel in Fig.\,\ref{hden_candidates}). This analysis suggests that an H/He composition of $\sim$\,99.5/0.5 is more probable (upper middle panel in Fig.\,\ref{hden_candidates}).
It is worth mentioning that using the XUV flux from the X-exoplanets model by \cite{Sanz_Forcada_2011}, which is about a factor of 3 smaller than that used here, we obtained a good agreement with an H/He of 98/2. That is to say, including the effects of several small flares in the stellar model (see Sect.\,\ref{fluxes}), we obtained a more ionised atmosphere and then a higher H/He.

In the case of \gj34, \cite{Salz2016} calculated the H$^0$ density for this planet, but they could not verify it because of the lack of \lya\ absorption measurements. 
More recently, however, \cite{Bourrier2018} measured the \lya\ absorption and concluded that the \citeauthor{Salz2016} model underestimates the H$^0$ density.
As for \hdu18, we also performed \lya\ absorption calculations and found that the profile of \cite{Salz2016} actually fit the observations of \cite{Bourrier2018} rather well. Further, we found that the errors in those observations are covered by the H$^0$ profile of \cite{Salz2016} when divided by $\sqrt{10}$ and multiplied by 10 (see lower panels in Fig.\,\ref{hden_candidates}).

Our $\chi^2$ analysis for \gj34 shows that the H$^0$ density profiles obtained with H/He ratios of 90/10 and 95/5 are significantly lower than those of \cite{Salz2018} at all altitudes (bottom left panel in Fig.\,\ref{hden_candidates}). They agree rather well, however, within the estimated uncertainties, for H/He ratios of 98/2 and 99/1 (bottom middle panel in Fig.\,\ref{hden_candidates}). For H/He ratios larger than 99.5/0.5, most of the H$^0$ density profiles that fit the \het\ absorption fall outside the estimated uncertainties of the H$^0$ \lya\ profile and hence are rather unlikely (bottom right panel in Fig.\,\ref{hden_candidates}).

%%%%%%%%%%%%%%%%%%%%%%%%%%%%%%%%
\subsection{MCMC analysis}
\label{mcmc}
%%%%%%%%%%%%%%%%%%%%%%%%%%%%%%%%
\begin{figure*}
\includegraphics[width=1\textwidth]{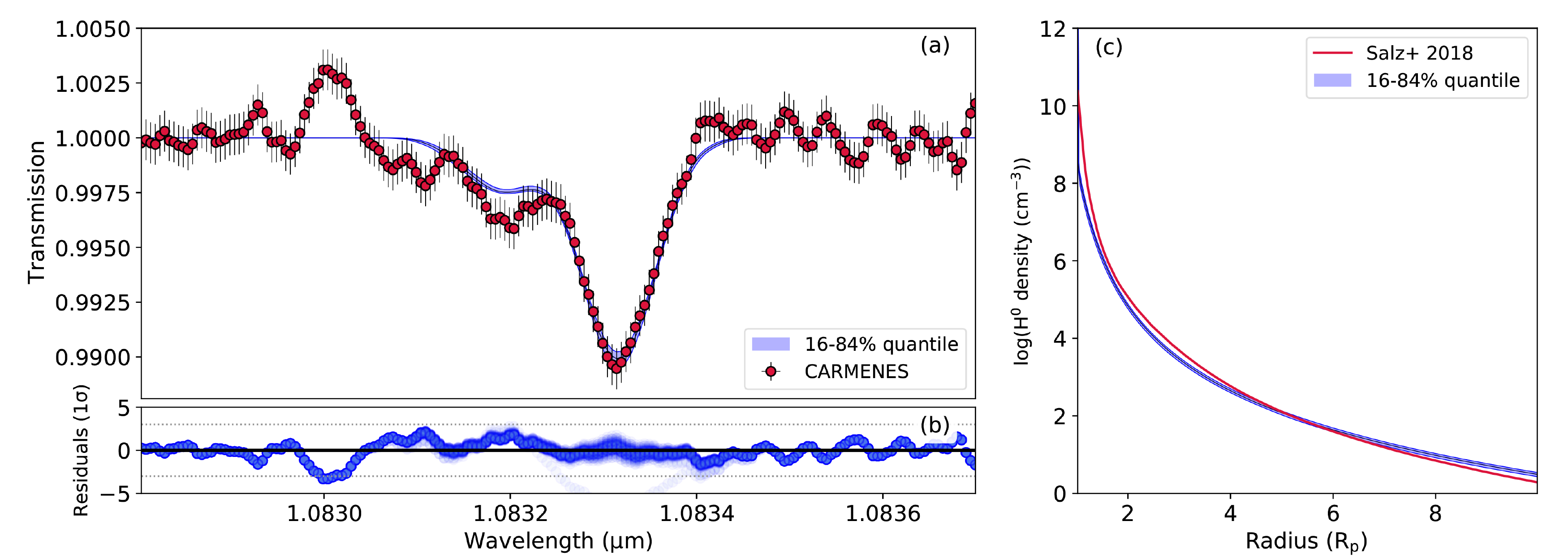} 
\includegraphics[width=1\textwidth]{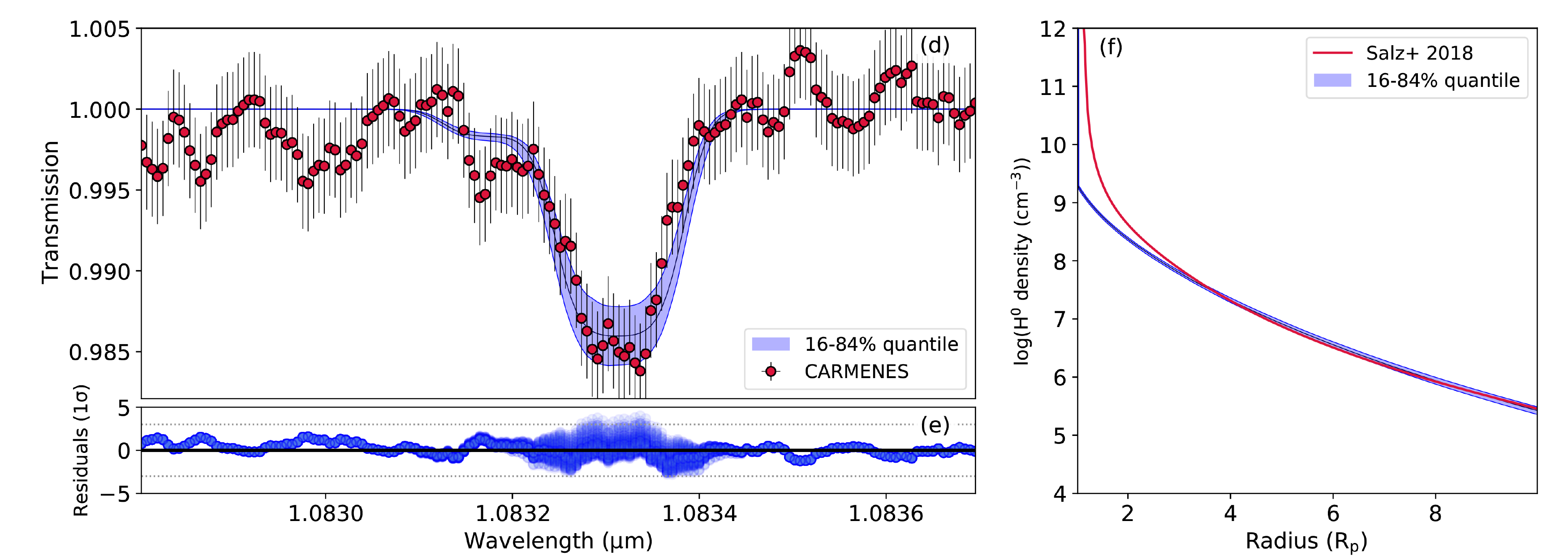} 
\caption{Bayesian inference of upper thermospheric conditions for \hdu18 and \gj34. (a) Comparison of the best-fit model spectra (blue shaded area) and of the measured He triplet line for \hdu18 (red data points). (b) Residual of the fitted models for \hdu18. The dotted horizontal lines mark 3-$\sigma$. (c) Comparison of the best-fit model H$^0$ density (shaded blue area) and of the estimated H$^0$ density from \lya\ measurements for \hdu18 (red data points). For the models, the blue area corresponds to the region of the posteriors between the 16 and 84\% quantiles. (d), (e), and (f) are similar to (a), (b), and (c), but for \gj34.}  
\label{mcmc_data}
\end{figure*}

\begin{figure}
\includegraphics[width=1\columnwidth]{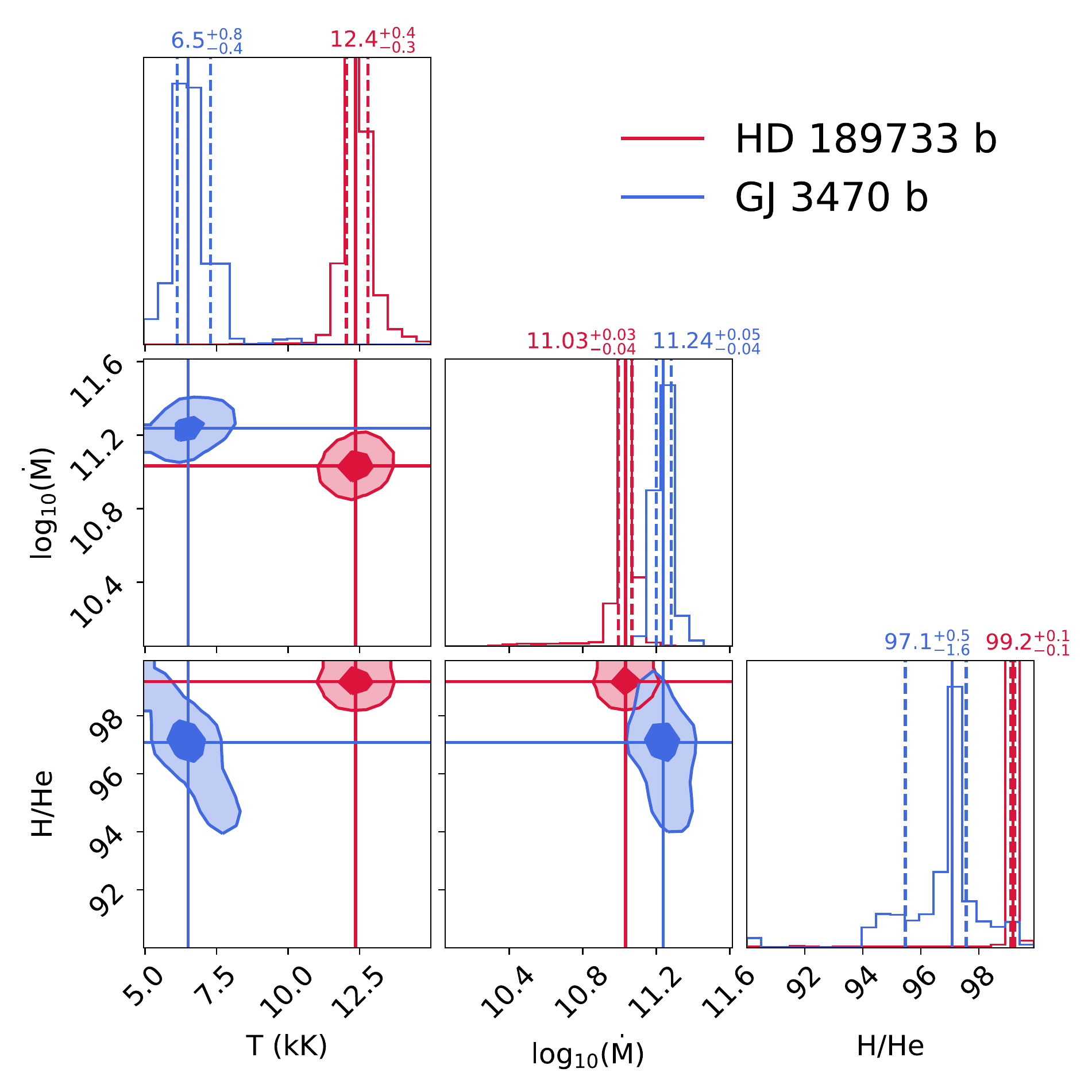}
\caption{Posterior probability distribution of our grid of models with respect to each of their parameter pairs as well as the marginalised distribution for each parameter for \hdu18 (red) and \gj34 (blue). The marginalised uncertainties are given as 16\% to 84\% quantiles. In the density maps, 1- and 2-$\sigma$ are given as 1-e$^{-0.5}$$\sim$0.393 and 1-e$^{-2.0}$$\sim$0.865, respectively, as are common for multivariate MCMC results.}  
\label{mcmc_corner}
\end{figure}

In order to investigate the posterior probability distribution of model parameters for the described grid of models in Section~\ref{model_grid}, 
and to determine if the simultaneous MCMC fit constrains the parameters further by sampling the parameter-space effectively, we used a python implementation of {\tt emcee} \citep{2010CAMCS...5...65G,2013PASP..125..306F}, called {\tt MCKM}. The {\tt MCKM} handles any arbitrary number of model parameters over regular (Cartesian) or irregularly spaced parameters, as well as any arbitrary number of data points and their respective covariance matrices.

We simultaneously fitted the observed He spectra and the H$^0$ density derived from \lya\ observations to constrain the temperatures, mass-loss rates, and H/He ratios of \hdu18 and \gj34 (see Fig.~\ref{mcmc_data}). 
We initialised 1000 walkers with uniform priors and with their range being the same as in the $\chi^2$ method to avoid differences between the two methods from different priors. Overall, the best models are consistent with the observations and no systematic residual is noticeable,  except for the H$^0$ density in the case of \gj34, where our models underestimate the density at low altitudes, R$<$2\,\rp. 
This could be due to the fact that we fitted global H$^0$ density profiles (i.e. not at individual altitudes) and all of our modelled H$^0$ profiles are systematically smaller than the profile of \cite{Salz2016} at R$<$2\,\rp. Further, because of the extended atmosphere of this planet, the atmospheric absorption at small radii is comparatively small and therefore has a weaker weight in the fitting of the whole atmosphere.  Thus, both facts together could explain the underestimation. 

Figure~\ref{mcmc_corner} shows the corner plot of the retrieved thermospheric parameters of \hdu18 (red) and \gj34 (blue). From this analysis, we find that the effective temperature, mass-loss rate, and H/He of \hdu18 are $T$=12\,400$^{+400}_{-300}$\,K, 
\mlr\,=\,(1.1$\pm0.1$)\,$\times$\,10$^{11}$\,\gs,
and H/He\,=\,(99.2/0.8)$\pm$0.1, respectively. The retrieved values for \gj34 are $T$=6500$^{+800}_{-400}$\,K,
\mlr\,=\,$1.74^{+0.21}_{-0.15}$\,$\times$\,$10^{11}$\,\gs,
and H/He=(97.1/2.9)$^{+0.5}_{-1.6}$.

%%%%%%%%%%%%%%%%%%%%%%%%%%%%%%%%
\subsection{Retrieved temperatures, mass-loss rates, and H/He composition}
\label{discussion}
%%%%%%%%%%%%%%%%%%%%%%%%%%%%%%%%

The results obtained by both the $\chi^2$ and the MCMC analysis are generally consistent. That is particularly clear in the case of \hdu18, for which the best fit of the H/He ratio derived from $\chi^2$ is very close to 99.5/0.5 and from the MCMC method we obtained H/He\,=\,(99.2/0.8)$\pm$0.1. The small uncertainty derived from the MCMC analysis is remarkable. Generally, one would expect a larger uncertainty range from the MCMC analysis than from the $\chi^2$ method, as the parameter space explored is wider. As explained above, a possible reason for the narrow MCMC posteriors might be that the MCMC fits the density profile at higher altitudes, but $\chi^2$ is fit at lower altitudes, particularly in the case of \gj34. The total number of data points at higher altitudes are more than that of the lower altitudes and hence a narrower posterior from such fit is not unexpected. The $\chi^2$ analysis suggests that they might be underestimated. Overall, as the most likely H/He ratio derived from both methods are very similar, we adopt the MCMC results of $T$=12\,400$^{+400}_{-300}$\,K, that is, with an uncertainty of about 400\,K; a  mass-loss rate of 
\mlr\,=\,(1.1$\pm0.1$)\,$\times$\,10$^{11}$\,\gs,
which is very well constrained (see the almost flat T/\mlr\ curves for high H/He ratios in left panel of Fig.\,\ref{chi2}), and an 
H/He\,=\,(99.2/0.8)$\pm$0.1, with possibly slightly larger uncertainties.

For \gj34, we have seen above that the MCMC analysis systematically underestimates the H$^0$ density at low altitudes. This could be the reason of the significantly smaller derived H/He ratio by this method, H/He=(97.1/2.9)$^{+0.5}_{-1.6}$, than that suggested by the $\chi^2$ analysis of 99/1 (in the range of 98/2--99.5/0.5; bottom panels in Fig.\,\ref{hden_candidates}). The general trend is to obtain larger H$^0$ densities for larger H/He ratios. The H/He ratio has a rather important impact on the mass-loss rate of this planet, as well as on the temperature (see the rather steep T/\mlr\ curves in the right panel of Fig.\,\ref{chi2}). Hence, the derived H/He has a significant impact on the resulting temperature and mass-loss rate ranges. Given the systematic underestimation of the H$^0$ density by MCMC, we are more inclined to adopt the H/He values and uncertainties derived from the $\chi^2$ method, which also embraces the high probability peak value at H/He=97 of MCMC (see Fig.\,\ref{mcmc_corner}). Thus, we conclude with a 
$T$\,=\,5100$\pm900$\,K, an 
\mlr\,=\,(1.9$\pm$1.1)$\times$10$^{11}$\,\gs, 
and an H/He=(98.5/1.5)$^{+1.0}_{-1.5}$ for \gj34.

It is interesting to note that of the three exoplanets undergoing hydrodynamic escape which have been analysed thus from their \het\ and \lya\ observations
(\hd20\ by \cite{Lampon2020}, and \hdu18\ and \gj34\ in this work), 
all show higher H/He ratios than the widely assumed 90/10, despite having rather different bulk parameters. 
Hence, this work suggests that an enrichment of H over He 
seems to be common in the upper atmospheres of giant exoplanets undergoing hydrodynamic escape.

One possibility to explain our results is that the escape of these atmospheres  originates above the homopause, where, due to diffusive separation, the atmosphere is enriched in H over the heavier He atoms \cite[see, e.g. Fig.\,14 in][]{Moses2005}. Our results, however, could also be consistent with an origin of the escape near the homopause, at least for \gj34, as \cite{Hu_2015} have shown that such enrichment is possible in Neptune and sub-Neptune exoplanets if the H mass-loss rate is comparable to its diffusion-limited mass-loss rate. This escape of H-enriched gas  can lead, in the case of Neptune and sub-Neptunes, to an He and metal enrichment in the lower atmosphere with further consequences on their composition (e.g. abundances of carbon and oxygen-bearing species) \citep{Hu_2015,Malsky_2020}. Furthermore, according to these authors, it may even change the mass-radius relationship of the planet.
In addition, our results are consistent with the depletion of atmospheric He by other processes as well, for example, with the formation of an H–He immiscibility layer in the interior of giant planets, as it produces He sequestration from the upper atmosphere \citep{Salpeter1973,Stevenson1975,Stevenson1980, Wilson2010}.
This result also suggests that the H/He ratio might play a more important role than expected in the detectability of \het. 
In addition to the spectral shape and intensity of the XUV stellar irradiation \cite[see, e.g.][]{Oklopcic_2019}, high H/He ratios could help to explain the non-detection of the \het\ in some highly irradiated exoplanets, such as GJ\,1214\,b and GJ\,9827\,d \citep[as suggested by][]{Kasper2020} as well as K2-100b \citep{Gaidos_2020}.

%%%%%%%%%%%%%%%%%%%%%%%%%%%%%%%%
\section{Comparison of temperatures and mass-loss rates with previous works }
\label{comparison_prev}
%%%%%%%%%%%%%%%%%%%%%%%%%%%%%%%%

%%%%%%%%%%%%%%%%%%%%%%%%%%%%%%%%
\subsection{\hdu18}
\label{comparison_HD189_he3}
%%%%%%%%%%%%%%%%%%%%%%%%%%%%%%%

\cite{guilluy_2020} observed the \het\ absorption with the GIANO-B high-resolution  spectrograph at the Telescopio Nazionale Galileo. They measured an \het\ mid-transit absorption of 0.75$\pm$0.03\%, slightly lower than the 0.88$\pm$0.04\% by \cite{Salz2018}; this was possibly due to the lower resolving power of GIANO-B with R $\approx$50\,000 compared to CARMENES with R$\approx$80\,400,  as argued by \cite{guilluy_2020}. 
These authors analysed their measurements with a 1D hydrodynamic isothermal Parker wind model \citep{Parker1958} for the thermosphere, 
and with a 3D particle model \citep{Bourrier_2013} above 1.2\,\rp, the altitude where they estimated the gas becomes non-collisional.
We applied a hydrodynamic model for the whole upper atmosphere because, as shown in our calculations, the altitude where the gas becomes non-collisional occurs far beyond the Roche lobe \cite[in agreement with][]{Salz2016,Odert_2020}.
They estimated a $T$--$\dot M$ relationship that favoured a thermosphere with a $T\approx$12\,000\,K and an \het\ density of 70 atoms\,cm$^{-3}$
at 1.2\,\rp\, assuming solar-like H/He composition. 
Despite the different modelling and assumptions between the two analyses, our derived \het\ distribution for the case of H/He\,=\,90/10 and $T$=\,12\,000\,K (see Fig.\,\ref{he3}) agrees well with their \het\ density at 1.2\,\rp. However, as we have shown in Sect.\,\ref{discussion}, our comparison with the \lya\ measurements suggests an H/He composition of $\sim$99.2/0.8, which leads to a higher mass-loss rate, $\dot M$\,$\approx$10$\times$10$^{10}$\,\gs (instead of $\dot M$\,=\,0.4$\times$10$^{10}$\,\gs\ for H/He=90/10),  for $T$=\,12\,000\,K (see Fig.\,\ref{chi2}). 
 
\cite{Salz2016} simulated the upper atmosphere of \hdu18 assuming that it is  composed by \textbf{$\sim$}90/10 of H/He, and using an incident $F_{\rm XUV}$\,$\approx$\,2.09$\times$10$^{4}$\,\ecs. They used a comprehensive 1D hydrodynamic model and fitted the \lya\ measurements of \cite{Lecavelier_des_Etangs_2012}. The resulting maximum temperature was 11\,800\,K and the mass-loss rate was $\dot M$\,$\approx$1.7$\times$10$^{10}$\,\gs. 
For the same H/He composition, mass-loss rate, and maximum temperature, we overestimated the measured \het\ absorption. 
Our results,
$\dot M$=(1.1$\pm0.1$)\,$\times$\,10$^{11}$\,\gs and $T$=12\,400$^{+400}_{-300}$\,K, show that our mass-loss rate range is larger by a factor of $\sim$6.3 
than that of \citeauthor{Salz2016}, while the maximum temperature is in good agreement. We note, however, that our stellar flux (see Table\,\ref{eqw}) is larger by a factor of 3 than that used by \citeauthor{Salz2016}, and our derived H/He composition is $\approx$ 99.2/0.8. 

\cite{Odert_2020} modelled the upper atmosphere of \hdu18y also using a hydrodynamic approach for fitting the \lya\ measurements of \cite{Lecavelier_des_Etangs_2012}. They assumed a thermosphere composed of H only and irradiated by $F_{\rm XUV}$\,$\approx$\,1.8\,$\times$10$^{4}$\,\ecs. They obtained a maximum temperature of 11\,000\,K and a mass-loss rate of 5.4$\times$10$^{10}$\,\gs.
We can see that our mass-loss rate range, $\dot M$=(1.1$\pm0.1$)\,$\times$\,10$^{11}$\,\gs, is larger by a factor of $\sim$2; our derived H/He composition is $\approx$\,99.2/0.8, 
which is close to the H-only composition assumed by \citeauthor{Odert_2020}; and our temperature range,
$T$=12\,400$^{+400}_{-300}$\,K, is slightly larger than that obtained by \citeauthor{Odert_2020} Therefore, considering that our F$_{XUV}$ is larger by a factor of 3, our results are in good agreement with those of \citeauthor{Odert_2020}

Overall, we obtain similar mass-loss rates as \cite{guilluy_2020}, when considering only the \het\ absorption, but larger rates by a factor of $\sim$\,25 
when also considering the \lya\ absorption. \cite{Salz2016} and \cite{Odert_2020} derived mass-loss rates from \lya\ measurements only. Our evaporation rates (constrained by both \het\ and \lya\ measurements) are larger (by a factor $\sim$6.3) than those of \citeauthor{Salz2016}, who assumed an H/He composition of
$\sim$\,90/10; however, they are in good agreement (slightly larger by a factor of $\sim$2) with those of \citeauthor{Odert_2020}, who assumed an H-only atmosphere.

%%%%%%%%%%%%%%%%%%%%%%%%%%%%%%%%
\subsection{ \gj34 }
\label{comparison_gj34}
%%%%%%%%%%%%%%%%%%%%%%%%%%%%%%%%

\cite{Ninan_2020} observed the \het\ absorption with the Habitable Zone Planet Finder near-infrared spectrograph (HPF), on the 10 m Hobby–Eberly Telescope at the McDonald Observatory. They measured an equivalent width of the \het\ mid-transit absorption of 0.012$\pm$0.002\,\AA, which is lower than that  calculated here of 0.0207\,\AA\ from the observations of \cite{Palle2020}. We note that differences could come from the different resolution power, as HPF has R$\approx$ 55\,000, 
or from the different spectral integration interval (we include the absorption of the weak line, i.e. from 10831.0 to 10834.5\,\AA). 
They found that the model by \cite{Salz2016} overestimates their \het\ measurements for this exoplanet, which agrees with our results.

\cite{Bourrier2018} modelled the upper atmosphere of \gj34\ with a parametrised thermosphere and a 3D particle model for non-collisional altitudes, assuming a thermospheric solar-like composition of H and He and a temperature of 7\,000\,K. They fitted their \lya\ observations obtaining an H$^0$ mass-loss rate of $\approx$1.5$\times$10$^{10}$\,\gs. Our derived total (i.e. all neutral, ionised, and excited species) mass-loss rate is
$\dot M$=(1.9$\pm1.1$)\,$\times$10$^{11}$\,\gs, 
which agrees with the lower limit imposed by their H$^0$ mass-loss rate.

\cite{Salz2016} modelled the upper atmosphere of \gj34\ with a comprehensive hydrodynamic model. For an H/He composition of $\sim$\,90/10 and a $F_{\rm XUV}\,\approx\,$ 7.8\,$\times$10$^{3}$\,\ecs, they estimated a $\dot M$\,$\approx$\,1.9$\times$10$^{11}$\,\gs\ and a maximum thermospheric temperature of $\approx$8600\,K. For the same H/He composition, mass-loss rate, and maximum temperature, we overestimated the \het\ absorption, which agrees with \cite{Ninan_2020}. Our results, 
$\dot M$=(1.9$\pm1.1$)\,$\times$10$^{11}$\,\gs, $T$=5100$\pm900$\,K, and H/He=(98.5/1.5)$^{+1.0}_{-1.5}$,
show that our mass-loss rates agree with those derived by \citeauthor{Salz2016}, although the temperature range is lower. We note, however, that our $F_{\rm XUV}$ is about a factor of 2 lower (see Table\,\ref{eqw}).

Overall, by using the same solar H/He ratio, mass-loss rate, and temperature as \cite{Salz2016}, we overestimated the \het\ measurements of  CARMENES, which agrees with the analysis of \cite{Ninan_2020} from their HPF measurements. 
Our mass-loss rates, constrained by \het\, and \lya\, observations, agree with those of \citeauthor{Salz2016}, but for a higher H/He composition 
(98.5/1.5)$^{+1.0}_{-1.5}$.
Also, our results agree with the lower limit of the mass-loss rate derived by \cite{Bourrier2018} from their \lya\ measurements.

%%%%%%%%%%%%%%%%%%%%%%%%%%%%%%%%
\section{Summary} \label{conclusions}
%%%%%%%%%%%%%%%%%%%%%%%%%%%%%%%%

In this work we have studied the hydrodynamic atmospheric escape of the hot Jupiter \hdu18 and the warm Neptune \gj34 by analysing the mid-transit \het\ absorption measurements observed with CARMENES \citep{Salz2018,Palle2020}. We used a 1D hydrodynamic model with spherical symmetry
(assuming substellar conditions apply to the whole planetary surface)
and a non-LTE model for computing the population of \het. 
As a further constraint, we also used the neutral
hydrogen density derived from \lya\ measurements in previous studies. 

The analysis of \hdu18 shows that the lower boundary conditions are very important in explaining the anomalously large absorption in the weaker \het\ line, which is caused by the hot and rather compressed upper atmosphere of this planet. It is worth mentioning that the absorption ratio of the weaker to stronger \het\ lines helps in constraining the mass-loss rate and lower boundary conditions of its atmosphere. Thus, spectrographs  with  sufficient resolution for discriminating the weak and strong lines provide further information about the evaporating planets.

The radial velocities of our hydrodynamic model for \hdu18 are too low to explain the  broad absorption profile of this planet. In order to fit it, we need to incorporate 
blue-shifted components at $-$3.5\,\kms\ and --11.5\,\kms\ covering nearly half and a quarter of the atmosphere's terminator, respectively, as well as a red component at 5.5\,\kms\ with 28\% of the terminator coverage. We also found that a thermospheric constant radial velocity of 40\,\kms, as derived by \cite{Seidel2020}, substantially overestimates the width of the \het\ lines, suggesting that such a large velocity is unlikely. 

In the case of \gj34, however, with a lower gravitational potential, our hydrodynamic model predicts gas radial velocities large enough to explain the width of the \het\ absorption profile very well. This, in fact, helps in constraining its mass-loss rate and temperature. Furthermore, the measured absorption profile exhibits a net blue shift at $-$3.2\,\kms, which we can be explained either by a net blue wind of the whole atmosphere at that velocity or by a combined atmosphere with a null blue shift below the Roche lobe and expanding at $-$5\,\kms\ above.

These two planets have a similar \het\ absorption, but rather different bulk parameters (see Table\,\ref{eqw}). In particular, the gravitational potential of \hdu18 is near a factor of 10 larger and \gj34 is irradiated in the XUV at about a factor of 14 smaller. In consequence, the characteristics of their upper atmospheres also differ significantly.
Thus, while \hdu18 has a rather compressed and warm atmosphere (12\,400$^{+400}_{-300}$\,K) with small gas radial velocities, \gj34 exhibits a very extended and cooler (5100$\pm900$\,K) atmosphere with large radial velocities. 
Also, while the upper atmosphere of \hdu18 is almost fully ionised beyond $\approx$1.1\rp, \gj34
exhibits a very wide ionisation front (from $\sim$1.25\,\rp\ to far beyond its Roche lobe). 
Overall, the gravitational potential and the irradiation balance result in comparable mass-loss rates, $\dot M$=(1.1$\pm0.1$)\,$\times$\,10$^{11}$\,\gs\ versus $\dot M$=(1.9$\pm1.1$)\,$\times$\,10$^{11}$\,\gs\ for \hdu18 and \gj34, respectively. The very different characteristics of these objects make them very suitable archetypes for benchmark studies on atmospheric loss.

We have further found that both planets have upper atmospheres with very low mean molecular masses (H/He=97/3--99.5/0.5). It is remarkable that the three exoplanets with evaporating atmospheres that have been studied so far by using both \het\ and \lya\ observations
(\hd20, \hdu18,\ and \gj34) all show higher H/He ratios than the commonly assumed 90/10, despite having very different bulk parameters. 
This H-enrichment of the upper atmospheres could be explained, on the one hand, by the escape originating above the homopause, where the atmosphere is expected to be depleted in He by diffusive separation. Another possibility, particularly in the case of the warm Neptune \gj34, is that the escape originates from deeper altitudes, around the homopause, according to the prediction of \cite{Hu_2015} of an H-enriched upper atmosphere for Neptune and sub-Neptunes undergoing hydrodynamic escape.
They further predict that it could lead to an He-enrichment of the lower atmosphere, with important consequences on their atmospheric composition (abundances of carbon and oxygen-bearing species) and even on their mass-radius relationship.
Our results also suggest that the H/He ratio might play a more important role than expected in the detectability of \het. The confirmation of this important result definitely calls for the study of other escaping atmospheres with concomitant \het\ and \lya\ measurements and for performing an independent analysis.

Here, we have analysed the \het\ mid-transit absorption spectra of \hdu18 and \gj34 with a 1D spherical model. However, comprehensive multi-fluid magneto-hydrodynamic 3D models are needed to provide more detailed information about the spatial and velocity distribution of the gas, the origin of the non-radial winds, and the influence of other processes (e.g. stellar wind, radiation pressure, or magnetic field interactions). In particular, the analysis of the ingress and egress spectra of these planets  would be very valuable.

\begin{acknowledgements}
We thank the referee for very useful comments. CARMENES is an instrument for the Centro Astron\'omico Hispano-Alem\'an (CAHA) at Calar Alto (Almer\'{\i}a, Spain), operated jointly by the Junta de Andaluc\'ia and the Instituto de Astrof\'isica de Andaluc\'ia (CSIC).
CARMENES was funded by the Max-Planck-Gesellschaft (MPG), the Consejo Superior de Investigaciones Cient\'{\i}ficas (CSIC), the Ministerio de Econom\'ia y Competitividad (MINECO) and the European Regional Development Fund (ERDF) through projects FICTS-2011-02, ICTS-2017-07-CAHA-4, and CAHA16-CE-3978, and the members of the CARMENES Consortium (Max-Planck-Institut f\"ur Astronomie,
Instituto de Astrof\'{\i}sica de Andaluc\'{\i}a,
Landessternwarte K\"onigstuhl, 
Institut de Ci\`encies de l'Espai, 
Institut f\"ur Astrophysik G\"ottingen, 
Universidad Complutense de Madrid, 
Th\"uringer Landessternwarte Tautenburg, 
Instituto de Astrof\'{\i}sica de Canarias, 
Hamburger Sternwarte, 
Centro de Astrobiolog\'{\i}a and 
Centro Astron\'omico Hispano-Alem\'an), 
with additional contributions by the MINECO, the Deutsche Forschungsgemeinschaft through the Major Research Instrumentation Programme and Research Unit FOR2544 ``Blue Planets around Red Stars'', the Klaus Tschira Stiftung, the states of Baden-W\"urttemberg and Niedersachsen, and by the Junta de Andaluc\'{\i}a.
We acknowledge financial support from the Agencia Estatal de Investigaci\'on of the Ministerio de Ciencia, Innovaci\'on y Universidades and the ERDF  through projects
ESP2016--76076--R,
ESP2017--87143--R,
PID2019-110689RB-I00/AEI/10.13039/501100011033,
BES--2015--074542,
PGC2018-099425--B--I00,
PID2019-109522GB-C51/2/3/4,     % CAB+IAA+IAC+UCM
PGC2018-098153-B-C33,           % ICE
AYA2016-79425-C3-1/2/3-P,       % UCM+CAB+IAA (until January 2020)
ESP2016-80435-C2-1-R,           % ICE (until January 2020) 
and the Centre of Excellence ``Severo Ochoa'' and ``Mar\'ia de Maeztu'' awards to the Instituto de Astrof\'isica de Canarias (SEV-2015-0548), Instituto de Astrof\'isica de Andaluc\'ia (SEV-2017-0709), and Centro de Astrobiolog\'ia (MDM-2017-0737), and the Generalitat de Catalunya/CERCA programme.
T.H. acknowledges support from the European Research Council under the Horizon 2020 Framework Program via the ERC Advanced Grant Origins 832428.
A.S.L. acknowledges funding from the European Research Council under the European Union's Horizon 2020 research and innovation program under grant agreement No 694513.
\end{acknowledgements}

\bibliographystyle{aa} % Standard, para [xx]
\bibliography{ref}

%%%%%%%% Appendix %%%%%%%%%%%%%%%%
\begin{appendix}

%%%%%%%%%%%%%%%%%%%%%%%%%%%%%%%%%%%%%%%%%%%%%%%%%%%%%%%%%%%%%%%%
\section{Data for calculating the stellar flux of HD\,189733.} \label{ap:sed}
%%%%%%%%%%%%%%%%%%%%%%%%%%%%%%%%%%%%%%%%%%%%%%%%%%%%%%%%%%%%%%%%
Data used in the modelling of the stellar flux of HD\,189733 in Sect.\,\ref{fluxes}. The spectrum was downloaded from the Hubble Spectral Legacy Archive (HSLA); a sum of 42 HD\,189733 spectra were acquired with the COS/G130M grating. The exposures were taken in 2009, 2013, and 2017 (proposals IDs 11673, 12984, and 14767).

%
%----------------------------------  Table A1
\begin{table}[htbp]
\caption[]{{\it HST}/COS line fluxes of HD 189733$^a$.}\label{tab:cosfluxes} 
\tabcolsep 4 pt
\begin{small}
\begin{tabular}{lrccrcl}
\hline \hline
%--------------------------------------------------------------
Ion & $\lambda_{\rm model}$ & $\log T_{\rm max}$ & $F_{\rm obs}$ & $S/N$ & Ratio & Blends \\
\hline
%--------------------------------------------------------------
\ion{Ne}{v} & 1145.5959 & 5.5 & 7.09e-17 & 4.5 & -0.02 &  \\
\ion{Si}{iii} & 1206.5019 & 4.9 & 1.10e-14 & 39.9 & -0.75 &  \\
\ion{O}{v} & 1218.3440 & 5.5 & 2.19e-15 & 39.8 & 0.21 &  \\
\ion{N}{v} & 1238.8218 & 5.4 & 3.32e-15 & 25.9 & 0.05 &  \\
\ion{N}{v} & 1242.8042 & 5.4 & 1.58e-15 & 17.6 & 0.03 &  \\
\ion{S}{ii} & 1253.8130 & 4.6 & 1.56e-16 & 9.6 & 0.03 &  \\
\ion{S}{ii} & 1259.5210 & 4.6 & 2.70e-16 & 11.7 & 0.04 &  \\
\ion{Si}{ii} & 1260.4240 & 4.6 & 5.00e-16 & 15.2 & -0.78 &  \\
\ion{Si}{ii} & 1264.7400 & 4.5 & 1.95e-15 & 18.5 & 0.12 & \ion{Si}{ii}1265.0040 \\
\ion{Si}{iii} & 1296.7280 & 4.9 & 1.13e-16 & 6.6 & 0.07 &  \\
\ion{Si}{iii} & 1298.9480 & 4.9 & 4.36e-16 & 7.4 & 0.37 & \ion{Si}{iii}1298.8940 \\
\ion{Si}{iii} & 1303.3250 & 4.9 & 1.36e-16 & 7.9 & 0.06 &  \\
\ion{Si}{ii} & 1309.2770 & 4.6 & 6.49e-16 & 11.0 & 0.02 &  \\
\ion{C}{ii} & 1323.9080 & 4.8 & 9.90e-17 & 5.9 & -0.40 & \ion{C}{ii}1323.9540 \\
\ion{C}{ii} & 1334.5350 & 4.7 & 8.96e-15 & 33.6 & -0.31 &  \\
\ion{C}{ii} & 1335.7100 & 4.7 & 1.72e-14 & 42.8 & 0.16 & \ion{C}{ii}1335.6650 \\
\ion{Fe}{iii} & 1364.2950 & 4.6 & 1.53e-16 & 9.8 & 0.01 &  \\
\ion{O}{v} & 1371.2960 & 5.5 & 2.96e-16 & 11.2 & -0.02 &  \\
\ion{Si}{iv} & 1393.7552 & 5.0 & 6.91e-15 & 25.8 & -0.02 &  \\
\ion{O}{iv} & 1401.1570 & 5.3 & 4.52e-16 & 10.8 & -0.75 &  \\
\ion{Si}{iv} & 1402.7704 & 5.0 & 3.55e-15 & 20.1 & -0.01 &  \\
\ion{S}{iv} & 1406.0160 & 5.1 & 6.40e-17 & 5.8 & -0.48 &  \\
%--------------------------------------------------------------
\hline
\end{tabular}

{$^a$ Line fluxes (in erg cm$^{-2}$ s$^{-1}$) 
  measured in {\it HST}/COS HD 189733 spectra. log $T_{\rm max}$ (K) indicates the maximum temperature of formation of the line (unweighted by the
  EMD). `Ratio' is the log($F_{\mathrm {obs}}$/$F_{\mathrm {pred}}$) 
  of the line. 
  Blends amounting to more than 5\% of the total flux for each line are
  indicated.}
\end{small}
\end{table}

%------------------------------------------------- begin table
\begin{table}[htbp]
\caption{Emission measure distribution of HD 189733.}\label{tabemd}
\renewcommand{\arraystretch}{1.3}  
\begin{center}
\begin{small}
\begin{tabular}{cc|cc}
\hline \hline
       {log~$T$ (K)} & {EM (cm$^{-3}$)\tablefootmark{a}} & {log~$T$ (K)} &
       {EM (cm$^{-3}$)} \\
\hline
%----------------
4.0 & 51.40: & 5.0 & 50.00$^{+0.10}_{-0.40}$ \\
4.1 & 51.25: & 5.1 & 49.90$^{+0.10}_{-0.40}$ \\
4.2 & 51.10: & 5.2 & 49.70$^{+0.10}_{-0.40}$ \\
4.3 & 50.90$^{+0.20}_{-0.20}$ & 5.3 & 49.40$^{+0.20}_{-0.10}$ \\
4.4 & 50.80$^{+0.20}_{-0.20}$ & 5.4 & 49.05$^{+0.05}_{-0.05}$ \\
4.5 & 50.55$^{+0.05}_{-0.15}$ & 5.5 & 48.85$^{+0.10}_{-0.10}$ \\
4.6 & 50.30$^{+0.20}_{-0.20}$ & 5.6 & 48.70$^{+0.10}_{-0.10}$ \\
4.7 & 50.20$^{+0.20}_{-0.30}$ & 5.7 & 48.70: \\
4.8 & 50.15$^{+0.15}_{-0.05}$ & 5.8 & 48.80: \\
4.9 & 50.10$^{+0.10}_{-0.20}$ & 5.9 & 48.90: \\
%---------------
\hline
\end{tabular}
\end{small}
\end{center}
\tablefoot{
\tablefoottext{a}{Emission measure (EM=log $\int N_{\rm e} N_{\rm H}
  {\rm d}V$), where $N_{\rm e}$
  and $N_{\rm H}$ are electron and hydrogen densities, in
cm$^{-3}$. Errors provided are not independent
between the different temperatures, as explained in \citet{san03}.}}
\renewcommand{\arraystretch}{1}
\end{table}
%--------------------------------------------- end table

%------------------------------------------------- begin table
\begin{table}[htbp]
\caption{Transition region abundances of HD 189733 (solar
  units\tablefootmark{a}).}\label{tababund} 
\tabcolsep 2.4pt
\begin{center}
\begin{small}
 \begin{tabular}{lrccc}
\hline \hline
      & {FIP} &  \multicolumn{2}{c}{Solar photosphere} & HD 189733 \\
  {X} & (eV)  &  Ref.$^a$ & (AG89)& [X/H]  \\
\hline
%----------------
Si &  8.15 & 7.51 & (7.56) &  -0.57$\pm$0.15 \\
Fe$^b$ &  7.87 & 7.50 & (7.67) &      -0.34      \\
S  & 10.36 & 7.12 & (7.21) & -1.22$\pm$0.24 \\
C  & 11.26 & 8.43 & (8.56) & -0.87$\pm$0.29  \\
O  & 13.61 & 8.69 & (8.93) & -0.86$\pm$0.31  \\
N  & 14.53 & 7.83 & (8.05) & -0.58$\pm$0.07 \\
Ne & 21.56 & 7.93 & (8.09) & -0.95$\pm$0.23 \\
%----------------
\hline
\end{tabular}
\end{small}
\end{center}
\tablefoot{
\tablefoottext{a}{Solar photospheric abundances from \citet{asp09},
  adopted in this table, are expressed on a logarithmic scale. 
Several values have been
updated in the literature since \citet[AG89]{anders} and they have also been listed
for easier comparison.}
\tablefoottext{b}{Fe abundance fixed to the coronal value (resulting from the fit of {\it XMM-Newton} EPIC data).}}
\end{table}
%--------------------------------------------- end table

\clearpage

%%%%%%%%%%%%%%%%%%%%%%%%%%%%%%%%%%%%%%%%%%%%%%%%%%%%%%%%%%%%%%%%
\section{Results for the derived H/He compositions}\label{ap:otherHHe}
%%%%%%%%%%%%%%%%%%%%%%%%%%%%%%%%%%%%%%%%%%%%%%%%%%%%%%%%%%%%%%%%
\het\ density profiles and gas radial velocities for the derived H/He ratios of \hdu18\ and \gj34.

\begin{figure}[htbp]
\includegraphics[angle=0.0, width=1.0\columnwidth]{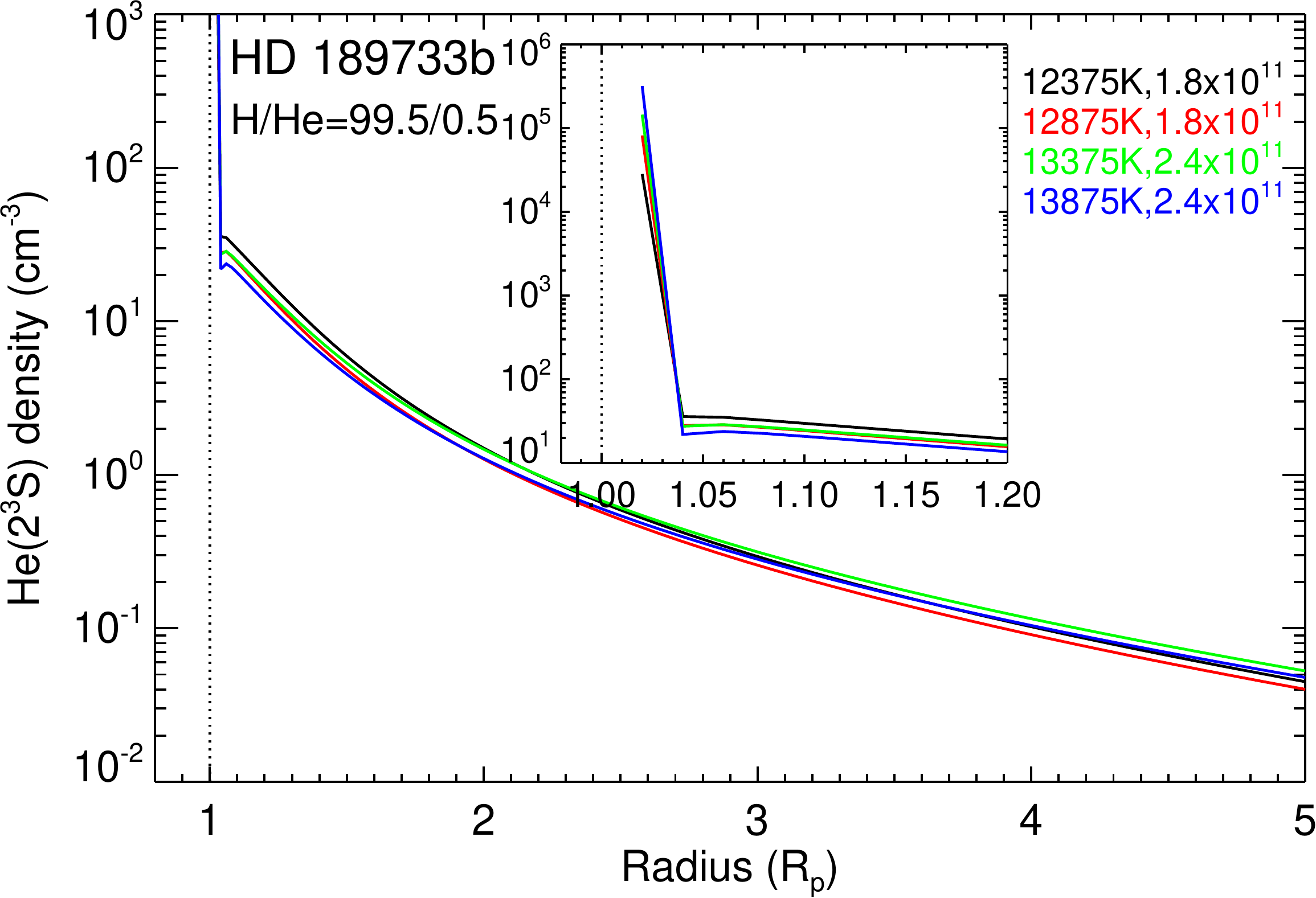}
\includegraphics[angle=0.0, width=1.0\columnwidth]{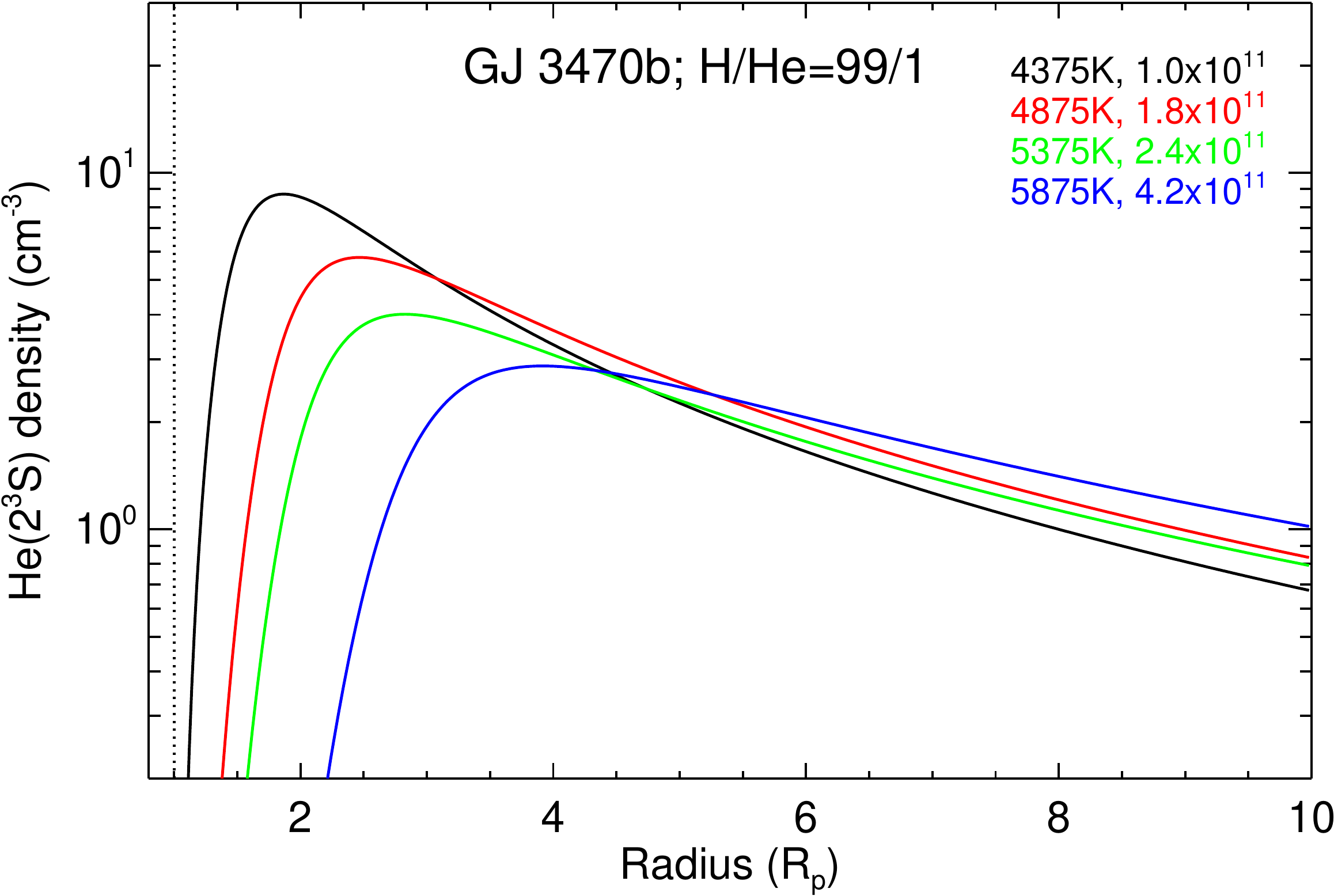}
\caption{\het\ concentration profiles that best fit the measured absorption (i.e. the filled circles in Fig.~\ref{chi2}) for \hdu18\ and an H/He ratio of 99.5/0.5 (upper panel), as well as for \gj34 and H/He=99/1 (lower panel).} \label{he3_supp} 
\end{figure}

\begin{figure}[htbp]
\includegraphics[angle=0.0, width=1.0\columnwidth]{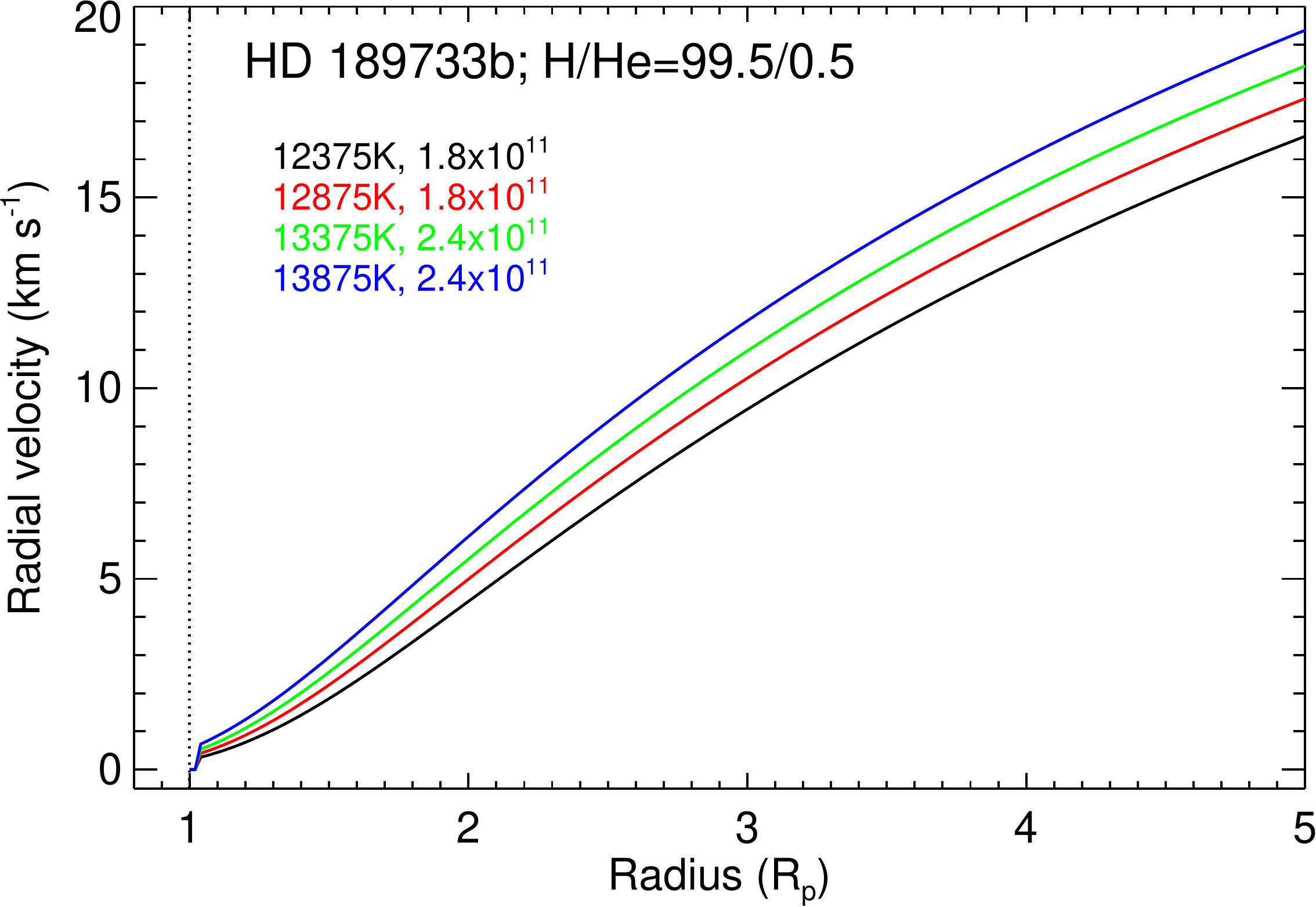}
\includegraphics[angle=0.0, width=1.0\columnwidth]{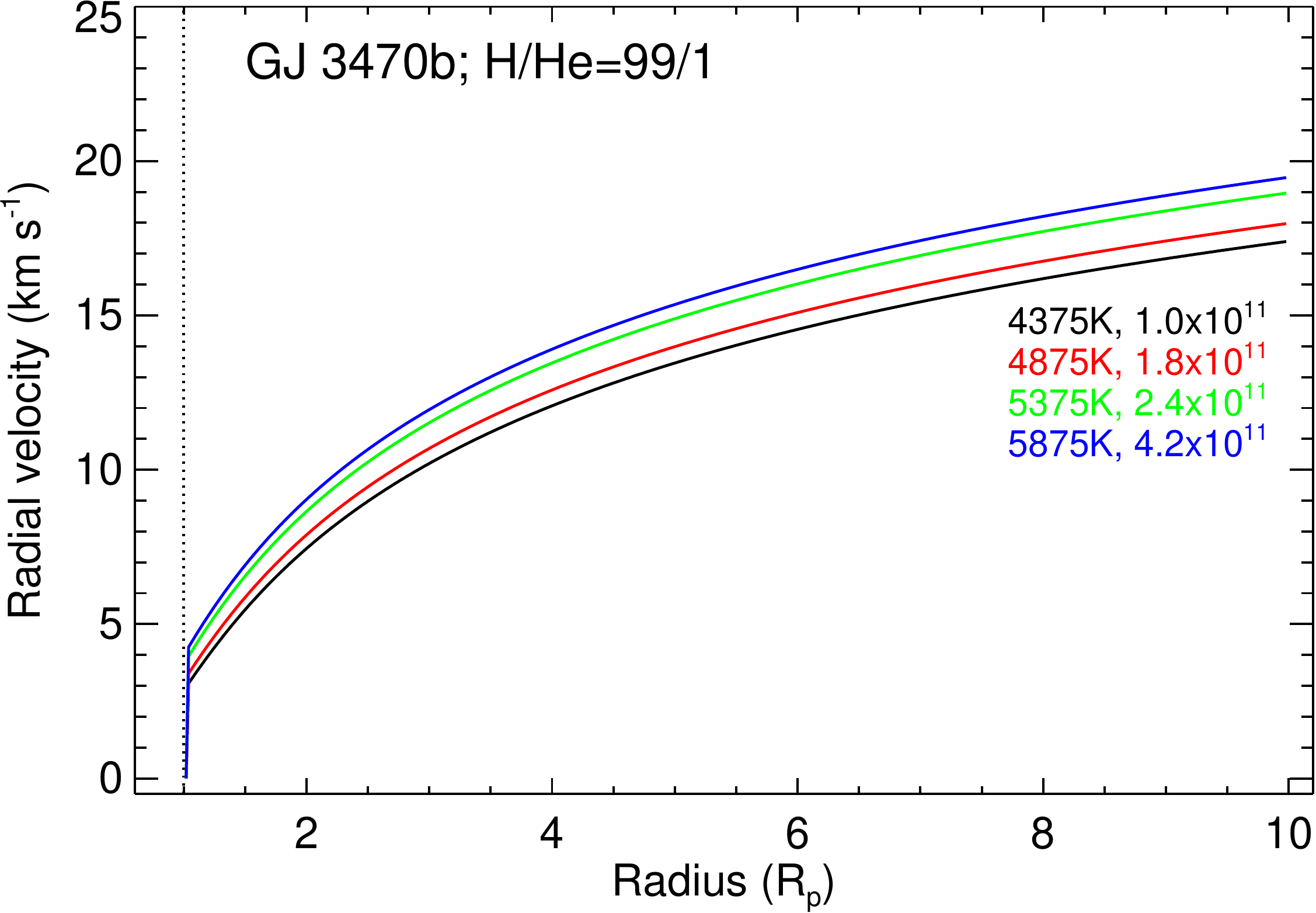}
\caption{Gas radial velocities of the hydrodynamic model for the best fit of the measured absorption (i.e. the filled circles in Fig.~\ref{chi2}) for \hdu18\ and an H/He ratio of 99.5/0.5 (upper panel), as well as for \gj34 and H/He=99/1 (lower panel).} \label{vel_supp} 
\end{figure}

%%%%%%%%%%%%%%%%%%%%%%%%%%%%%%%%%%%%%%%%%%%%%%%%%%%%%%%%%%%%%%%%
\section{Results for neglecting the broadening turbulence}\label{ap:chi2_noturb}
%%%%%%%%%%%%%%%%%%%%%%%%%%%%%%%%%%%%%%%%%%%%%%%%%%%%%%%%%%%%%%%%

Results for the $T$-\mlr\ map for \hdu18 when the turbulence broadening is not considered. 

\begin{figure}[htbp]
\includegraphics[angle=0.0, width=1.\columnwidth]{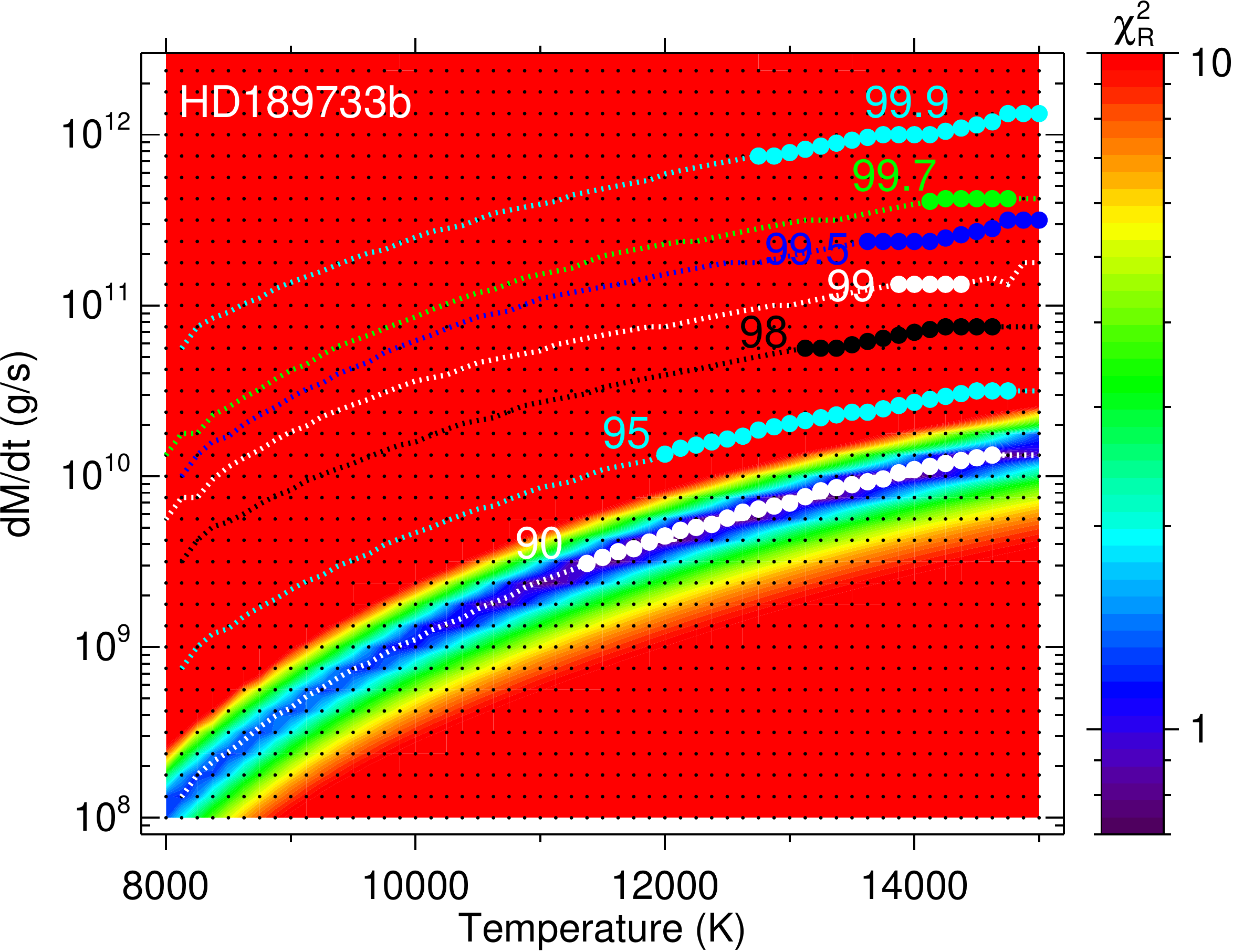}
\caption{Contour maps of the reduced $\chi^2$ of the model of the helium triplet absorption for \hdu18 for several H/He ratios (as in Fig.\,\ref{chi2}, left panel) when the turbulence broadening is not considered. The filled circles highlight the best fits (constrained ranges). The black dots represent the grid of the models.}  
\label{chi2_noturb}
\end{figure}

\end{appendix}

\end{document}